\documentclass[%
reprint,
superscriptaddress,
nofootinbib,
showkeys,
amsmath,amssymb,
aps,
]{revtex4-2}

\usepackage{graphicx}
\usepackage{dcolumn}
\usepackage{bm}
\usepackage[french,english]{babel}
\usepackage[T1]{fontenc} 
\usepackage{enumitem}
\setlist{nolistsep}
\usepackage{cancel}
\usepackage{siunitx}
\usepackage[normalem]{ulem}
\usepackage{xcolor}
\usepackage{float}
\usepackage{appendix}

\begin{document}

\preprint{APS/123-QED}



\title{\Large{Broadband Magnetless Isolation in a \\ Flux-Pumped, Dispersion-Engineered Transmission Line}
}

\author{M. Demarets}
\email{mael.demarets@imec.be}
\affiliation{imec, Kapeldreef 75, Leuven, Belgium} 
\affiliation{Department of Electrical Engineering (ESAT), KU Leuven, Leuven, Belgium}
\author{A. M. Vadiraj}
\email{vadirajrao@imec.be}
\affiliation{imec, Kapeldreef 75, Leuven, Belgium} 
\author{C. Caloz}
\affiliation{Department of Electrical Engineering (ESAT), KU Leuven, Leuven, Belgium}
\author{K. De Greve}
\affiliation{imec, Kapeldreef 75, Leuven, Belgium}
\affiliation{Department of Electrical Engineering (ESAT), KU Leuven, Leuven, Belgium}
\affiliation{Proximus Chair in Quantum Science and Technology, KU Leuven, Leuven, Belgium}

\date{\today}

\begin{abstract}
Isolators are commonly found in the amplification chain of microwave setups to shield sensitive devices such as superconducting qubits from noise and back-scattered signals. Conventional ferrite-based isolators are bulky, lossy and rely on strong magnetic fields, which pose challenges for their co-integration in large-scale superconducting devices. Although several magnetless approaches based on parametric modulation have been explored to overcome these limitations, none has yet experimentally demonstrated wideband isolation on par with ferrite devices.
Here, we propose a compact modulation-based isolator that achieves large isolation bandwidth using a dispersion-engineered transmission line. The engineered line forms an effective two-mode system that enables broadband isolation by supporting adiabatic mode conversion over a wide instantaneous bandwidth. Numerical simulations show that this architecture can provide more than $20\text{ dB}$ isolation across $4-8\text{ GHz}$, matching the performance of ferrite-based isolators. Moreover, we propose an on-chip superconducting device implementation that shows promise against parameter variations and enables a scalable path for co-integration with future large-scale superconducting systems.
\end{abstract}


\maketitle

\section{\label{sec:Intro}Introduction}

Nonreciprocal devices are common in communication and measurement systems. In particular, isolators are typically used to protect sensitive elements from noise or back-scattered signals. These isolators are found in various applications, across a broad range of frequencies, from microwave to optical, and in a broad range of operating conditions, from room temperature to the cryogenic regime. Conventional microwave isolators, based on ferrites biased by strong magnetic fields, provide high isolation over multi-gigahertz bandwidth \cite{Fay1965,Pozar2011}. As such, they have been widely adopted in ultra-low-noise environments -- for example in the amplification chain of superconducting quantum circuits \cite{Arute2019,Storz2023}. Despite their success, ferrite isolators pose significant challenges for large-scale integration because they are bulky, lossy and difficult to integrate on chip. Moreover, the strong magnetic fields required to break reciprocity in these isolators can disturb and deteriorate the performance of nearby circuitry, such as superconducting qubits, which further complicates their co-integration. \par

Many approaches have been proposed to achieve magnetless nonreciprocity \cite{Caloz2018,Kord2020,Nagulu2021}. Among them, integrated solutions based on parametric modulation have shown strong nonreciprocal transmission with minimal added noise and low power dissipation, making them particularly well-suited for cryogenic ultra-low-noise environments. Such cryogenic parametric devices have enabled magnetless isolation \cite{Barzanjeh2017,Chapman2017,Muller2018,Ranzani2017,Naghiloo2021},
directional amplification \cite{Yurke1996,Abdo2014,Yaakobi2013,Zorin2019,HoEom2012,Vissers2016,Chaudhuri2017,Malnou2021,Obrien2014,KowKamal2025,Chang2025,Wang2025,Macklin2015,Planat2020,Roudsari2023,Gaydamachenko2025}
and the simultaneous realization of both effects in the same structure \cite{Ranadive2024,Malnou2025,Kow2025,Abdo2013,Metelmann2015,Sliwa2015,Lecocq2017}.
However, to the best of our knowledge, no integrated magnetless device has yet been experimentally demonstrated with the large isolation bandwidth required for shielding and multiplexed readout as achieved with ferrite-based isolators.\par
To address this gap, we introduce an architecture that achieves broadband magnetless isolation within a single impedance-matched waveguide. Leveraging dispersion engineering \cite{HoEom2012,Obrien2014,Macklin2015,Vissers2016,Chaudhuri2017,Planat2020,Roudsari2023,Malnou2021,Chang2025,Gaydamachenko2025,Wang2025,Ranadive2024,Kow2025,Malnou2025,KowKamal2025}, the waveguide acts as an effective two-mode system with directional coupling and matched group velocities. This  effective two-mode system enables wideband magnetless isolation by supporting adiabatic mode conversion \cite{Naghiloo2021} over a broad frequency range (Sec.~\ref{sec:SystemDescription}). We present a compact microwave circuit implementation based on a $50 \ \Omega$ flux-modulated transmission line \cite{Zorin2019}, which provides direct access to the coupled modes (Sec.~\ref{sec:Implementation}). Simulations show that this implementation achieves
isolation, insertion loss, and bandwidth comparable to conventional ferrite isolators, while showing promise against parameter spread arising from fabrication-related challenges (Sec.~\ref{sec:Simulations}). With its compact implementation and state-of-the-art simulated performance, the proposed architecture could potentially replace conventional isolators and support the development of large-scale superconducting circuits (Sec.~\ref{sec:Discussion}).

\section{System Description} \label{sec:SystemDescription}
The proposed architecture relies on three key elements to achieve broadband magnetless microwave isolation: \textit{directional parametric coupling}, \textit{dispersion engineering} and \textit{adiabatic mode conversion}.
\subsection{Directional Parametric Coupling} \label{subsec:DirectionalParametricCoupling}
\begin{figure}[!ht]
\centering
\includegraphics[width=\linewidth]{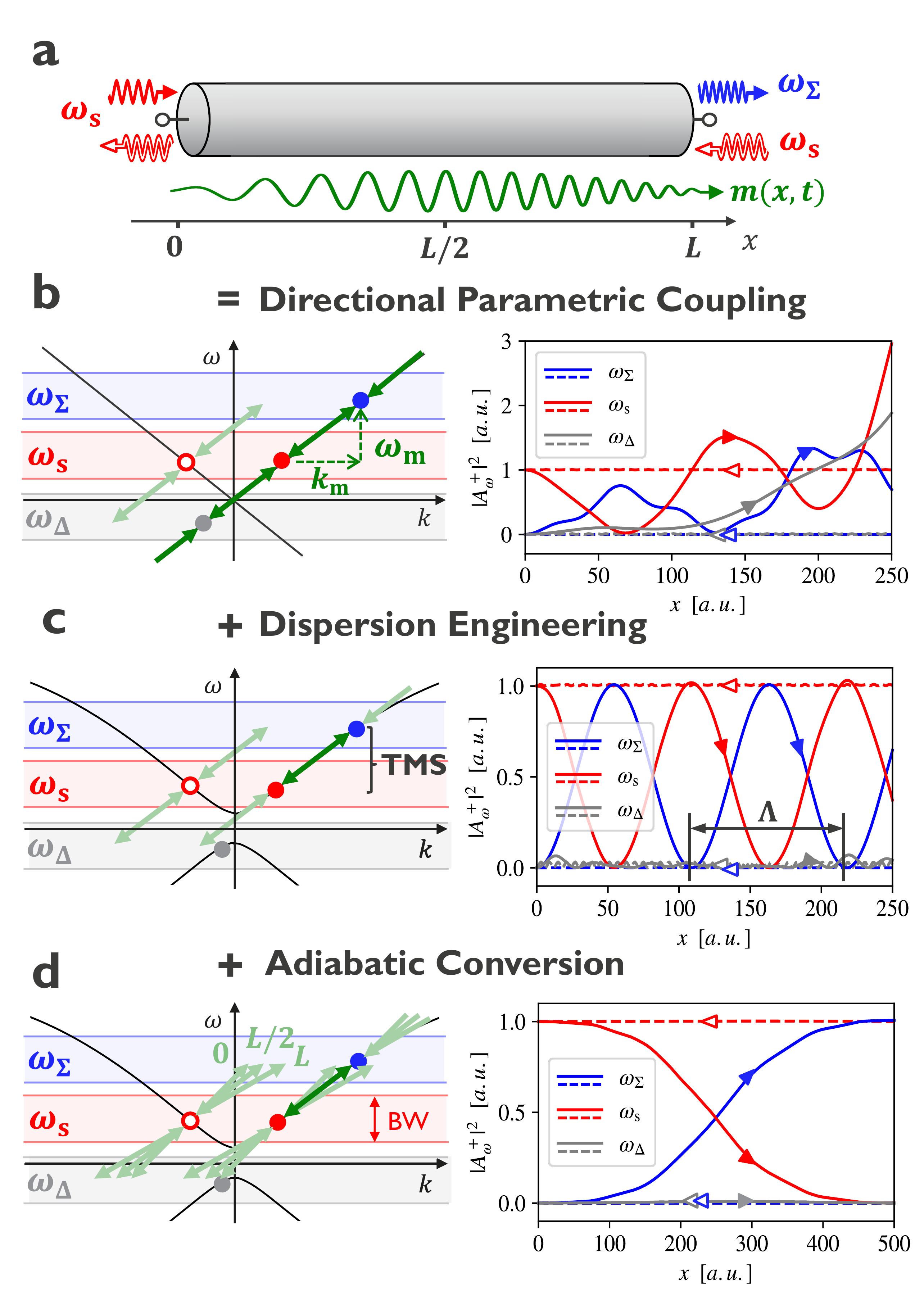}
\caption{\textbf{Broadband isolation in a modulated waveguide} \textbf{(a)} Nonreciprocal frequency up-conversion realized with a propagating parametric modulation $m(x,t)$ whose wavenumber and amplitude are adiabatically varied in space.
\textbf{(b)} 
The propagating parametric modulation
breaks time-reversal symmetry and produces direction-dependent coupling (green arrow) between modes of the waveguide. When phase-matched (solid lines), the modulation produces both amplification and conversion with the modes at $\omega_\Delta$ and $\omega_\Sigma$, respectively. In the reverse direction (dashed lines), the modulation is not phase-matched and parametric coupling is strongly suppressed.
\textbf{(c)} At lower frequencies, dispersion engineering prevents parametric coupling to the modes $\omega_\Delta$ below the bandgap which suppresses the undesired amplification process. At higher frequencies, the engineered dispersion curve limits the coupling to higher modes so that only the two modes at $\omega_{\mathrm{s}}$ and $\omega_\Sigma$ are parametrically coupled. These two modes hence form an effective two-mode system (TMS) and periodically exchange photons with wavelength $\Lambda$.
\textbf{(d)} 
The adiabatic modulation is swept through the phase-matching conditions and converts the right-propagating signal (solid lines) in the bandwidth of interest (in red) to higher-frequency modes (in blue) without converting back which provides high isolation over the whole operating bandwidth ($\text{BW}$). In the reverse direction (dashed lines), parametric coupling is negligible, resulting in low insertion loss.
}
\label{fig:SystemDescription}
\end{figure}

Consider a two-port waveguide supporting the propagation of electromagnetic waves as illustrated in Fig.~\ref{fig:SystemDescription}a. The first element used is a propagating parametric modulation $m(x,t)$ of the waveguides' permittivity, permeability, or conductivity, which breaks the reciprocity of the waveguide and induces a direction-dependent coupling between modes at different frequencies \cite{Yaakobi2013,Zorin2019,HoEom2012,Obrien2014,Macklin2015,Vissers2016,Chaudhuri2017,Planat2020,Roudsari2023,Malnou2021,Chang2025,Gaydamachenko2025,Wang2025,Ranzani2017,Naghiloo2021,Ranadive2024,Kow2025,Malnou2025,KowKamal2025}.
When no modulation is applied, i.e. $m(x,t) = 0$, the normal modes of the waveguide are a set of uncoupled right- ($+$) and left-propagating ($-$) waves with envelopes $A^{\pm}_{\omega}$, where $|A^{\pm}_{\omega}|^2$ is the average number of photons propagating in that mode and where each mode satisfy the linear dispersion relation $\omega(k)$ plotted in Fig.~\ref{fig:SystemDescription}b. 
When a modulation is applied, i.e. $m(x,t)\neq0$, the right- and left-propagating modes are no longer the normal modes of the perturbed waveguide and coupling arises between left- and right-propagating waves. Due to this modulation-induced coupling, the envelopes $A^{\pm}_{\omega}(x)$ become slow-varying functions of position. \par

Consider a right-propagating modulation $m(x,t) = |m| \sin \left(\omega_{\mathrm{m}} t - \int_0^x k_{\mathrm{m}}(x') \ \mathrm{d}x' \right)$. The modulation, oscillating over time at angular frequency $\omega_{\mathrm{m}}$, creates parametric coupling between the modes at $\omega_{\mathrm{s}}$, $\omega_\Sigma = \omega_{\mathrm{s}} + \omega_{\mathrm{m}}$ and $\omega_\Delta = \omega_{\mathrm{s}} - \omega_{\mathrm{m}}$, where the non-energy conserving terms oscillating rapidly over time were neglected (Appendix~\ref{appendix:WaveEquations}). Moreover, the modulation also oscillates over space with angular frequency $k_{\mathrm{m}}(x)$, so the fast-oscillating terms in space can also be neglected, which decouples the right- and left-propagating modes and yields the following dynamics (Appendix~\ref{appendix:WaveEquations})
\begin{align}
\frac{\mathrm{d}A^{\pm}_{\omega_{\mathrm{s}}}}{\mathrm{d} x} & =
-\frac{|m|\sqrt{|\omega_{\mathrm{s}}|}}{\sin(k_{\mathrm{s}})\sqrt{Z_0(\omega_{\mathrm{s}})}} \label{eq:CoupledModes}\\
\times & \Bigg\{ \sin\left(\frac{k_{\Sigma}\mp k_{\mathrm{m}}}{2}\right) \sin\left(\frac{k_{\Sigma}}{2}\right) \frac{\sqrt{Z_0(\omega_{\Sigma})}}{\sqrt{|\omega_{\Sigma}|}} A^{\pm}_{\omega_{\Sigma}} e^{\pm i \Delta \theta^{\pm}_{\Sigma}} \nonumber \\
&-
\sin\left(\frac{k_{\Delta}\pm k_{\mathrm{m}}}{2}\right) \sin\left(\frac{k_{\Delta}}{2}\right) \frac{\sqrt{Z_0(\omega_{\Delta})}}{\sqrt{|\omega_{\Delta}|}} A^{\pm}_{\omega_{\Delta}}e^{\pm i \Delta \theta^{\pm}_{\Delta}} \Bigg\} \nonumber
\end{align}
with $Z_0(\omega)$ being the characteristic impedance of the waveguide, where we used $k_{\Sigma/\Delta} = k(\omega_{\Sigma/\Delta})$ for ease of notation and defined the accumulated phase mismatch
\begin{subequations}
  \label{eq:PhaseMismatch}
  \begin{equation}
    \Delta \theta^{\pm}_{\Sigma} = \int_0^x \Delta k^\pm_\Sigma\ \mathrm{d}x' = \int_0^x (k_{\Sigma}-k_{\mathrm{s}} \mp k_{\mathrm{m}}) \ \mathrm{d}x' 
  \end{equation}
  \begin{equation}
    \Delta \theta^{\pm}_{\Delta} = \int_0^x \Delta k^\pm_\Delta\ \mathrm{d}x' = \int_0^x (k_{\Delta}-k_{\mathrm{s}} \pm k_{\mathrm{m}}) \ \mathrm{d}x' 
  \end{equation}
\end{subequations}
As shown in Fig.~\ref{fig:SystemDescription}b, phase matching is satisfied when the temporal and spatial frequencies of the modulation match the differences in temporal and spatial frequencies between the two coupled modes, e.g. $\omega_{\mathrm{m}}=|\omega_{\Sigma}-\omega_{\mathrm{s}}|$ and $k_{\mathrm{m}}=|k_{\Sigma}-k_{\mathrm{s}}|$. As can be seen in Fig.~\ref{fig:SystemDescription}b, the same propagating modulation cannot be simultaneously phase-matched in both directions of propagation, i.e. for right- and left-propagating modes with positive and negative $k(\omega)$ values, respectively. Such modulation-induced parametric coupling is therefore a direction-dependent process \cite{Ranzani2017,Naghiloo2021,Yaakobi2013,Zorin2019,HoEom2012,Obrien2014,Macklin2015,Vissers2016,Chaudhuri2017,Planat2020,Roudsari2023,Malnou2021,Chang2025,Gaydamachenko2025,Wang2025,Ranadive2024,Kow2025,Malnou2025,KowKamal2025}.\par
Parametric coupling between propagating modes of the waveguide results in nonreciprocal parametric amplification or conversion depending on the frequencies of the coupled mode $\omega_i$, i.e. $\omega_{\mathrm{m}}=\omega_{\mathrm{s}}+\omega_i$ or $\omega_{\mathrm{m}}= |\omega_i-\omega_{\mathrm{s}}|$, respectively. As shown in Fig.~\ref{fig:SystemDescription}b, parametric amplification of the right-propagating modes at $\omega_{\mathrm{s}}$ and $\omega_\Delta$ causes the number of photons in the modes to grow as they propagate. On the other hand, parametric mode conversion results in energy exchange between the coupled right-propagating modes at $\omega_{\mathrm{s}}$, $\omega_\Sigma = \omega_{\mathrm{s}} + \omega_{\mathrm{m}}$, $\omega_{\mathrm{s}} + 2\omega_{\mathrm{m}}$, etc. In the reverse direction, the large phase-mismatch suppresses the parametric coupling and these effects are not visible for left-propagating modes (see Fig.~\ref{fig:SystemDescription}b). The modulated waveguide is thus nonreciprocal. 

\subsection{Dispersion Engineering}\label{subsec:DispEng}
As discussed in the previous section, parametric coupling between propagating modes of the waveguide can produce two distinct effects.
Firstly, coupling to the higher-frequency mode at $\omega_\Sigma=\omega_{\mathrm{s}} + \ \omega_{\mathrm{m}}$ enables coherent energy exchange with this mode and can be used to up-convert signals outside the bandwidth of interest to achieve high isolation.
In contrast, coupling to the lower frequency mode at $\omega_\Delta$ can lead to parametric amplification whenever $\omega_{\mathrm{s}}<\omega_{\mathrm{m}}$, as seen in Fig.~\ref{fig:SystemDescription}b. The amplification of signals in the bandwidth of interest undermines the depletion of the bandwidth and hence limits the achievable isolation. 
The second key element of this work is therefore the use of dispersion engineering \cite{HoEom2012,Obrien2014,Macklin2015,Vissers2016,Chaudhuri2017,Planat2020,Roudsari2023,Malnou2021,Chang2025,Gaydamachenko2025,Wang2025,Ranadive2024,Kow2025,Malnou2025,KowKamal2025},
at lower and higher frequencies to mitigate the spurious amplification process and realize an effective two-mode system with directional coupling. 

At lower frequencies, the dispersion relation is engineered so the parametric modulation is strongly phase-mismatched resulting in weak parametric coupling and suppressed amplification. For that purpose, a bandgap is opened in the dispersion curve at these lower frequencies, which results in a phase-mismatch $\Delta k^{+}_\Delta$ that limits the coupling to $\omega_\Delta$ and strongly suppresses the undesired amplification process, as seen in Fig.~\ref{fig:SystemDescription}c. 
At higher-frequencies, the dispersion curve is also engineered to prevent phase matching and conversion to higher-frequency modes at $\omega_\Sigma+\omega_{\mathrm{m}}$. With both ends of the dispersion curve engineered, the parametric coupling is then limited to only two modes at $\omega_{\mathrm{s}}$ and $\omega_{\Sigma}$ that periodically exchange energy, as seen in Fig.\ref{fig:SystemDescription}c.\par

Neglecting the modes that are strongly phase-mismatched, i.e. with $\Delta k^{\pm}\gg k$, the coupled mode equations of Eq.~\eqref{eq:CoupledModes} simplify to a two-mode system. In the rotating frame corresponding to the transformation ${A}^{+}_{\omega_{\Sigma}} \rightarrow A^{+}_{\omega_{\Sigma}} e^{i\Delta \theta_\Sigma^+(x)/2}$ and ${A}^{+}_{\omega_{\mathrm{s}}} \rightarrow A^{+}_{\omega_{\mathrm{s}}} e^{-i\Delta \theta_\Sigma^+(x)/2} $, this system of equations can be recast as
\begin{equation}\label{eq:CoupledTLSRotatingFrame}
\frac{\mathrm{d}}{\mathrm{d}x} \begin{bmatrix} A^{+}_{\omega_{\Sigma}} \\ A^{+}_{\omega_{\mathrm{s}}}
\end{bmatrix}
\approx \begin{bmatrix} \frac{i}{2}\Delta k^+_\Sigma & g \\
-g & -\frac{i}{2}\Delta k^+_\Sigma
\end{bmatrix}
\cdot 
\begin{bmatrix} A^{+}_{\omega_{\Sigma}} \\ A^{+}_{\omega_{\mathrm{s}}}
\end{bmatrix}
\end{equation}
with off-diagonal coupling rates (Appendix~\ref{appendix:WaveEquations})
\begin{equation}\label{eq:TLSCouplingRate}
g = |m| \frac{\sqrt{|\omega_\Sigma|}}{\sqrt{|\omega_{\mathrm{s}}|}}\frac{\sqrt{Z_0(\omega_{\mathrm{s}})}}{\sqrt{Z_0(\omega_{\Sigma})}}\frac{\sin\left(k_{\mathrm{s}}/2\right)}{\sin(k_{\Sigma})}\sin\left(\frac{k_{\mathrm{s}}-k_{\mathrm{p}}}{2}\right)
\end{equation}
The two coupled right-propagating modes hence exchange photons back and forth with a Rabi wavelength $\Lambda =2\pi/\sqrt{|g|^2+(\Delta k^+_\Sigma)^2/4} $, as observed in Fig.~\ref{fig:SystemDescription}c. Maximum isolation is obtained with an optimal $|g|$ value that leaves the right-propagating input mode at $\omega_{\mathrm{s}}$ fully depleted at the end of the line, that is $L/\Lambda = 1/2 \ (\text{mod }1)$ with $L$ the length of the line. The right-propagating signal is then fully transferred to the higher-frequency mode $\omega_\Sigma$ outside the bandwidth of interest, which can later be dissipated with absorptive filters \cite{Kedziora2025}.

\subsection{Adiabatic Mode Conversion}\label{subsec:AdiabModeConv}
After combining parametric modulation and dispersion engineering to create an effective two-mode system with directional coupling, the third key element in the proposed isolator is the use of \textit{adiabatic mode conversion} \cite{Naghiloo2021} to achieve broadband isolation. 
According to Eq.~\eqref{eq:TLSCouplingRate}, the coupling rate $|g|$ depends on the input signal frequency $\omega_{\mathrm{s}}$ and so the Rabi wavenumber $\Lambda$ also varies with $\omega_{\mathrm{s}}$. Consequently, full depletion of the input mode can happen only over the narrow ranges of frequencies where $L/\Lambda(\omega_{\mathrm{s}}) = 1/2 \ (\text{mod }1)$.
To overcome that bandwidth constraint, a quasi-adiabatic parametric modulation is used to leave the mode of interest fully depleted regardless of its frequency, as shown in Fig.~\ref{fig:SystemDescription}d, at an acceptable cost of device length increase. \par

\begin{figure}[!ht]
\centering
\includegraphics[width=\linewidth]{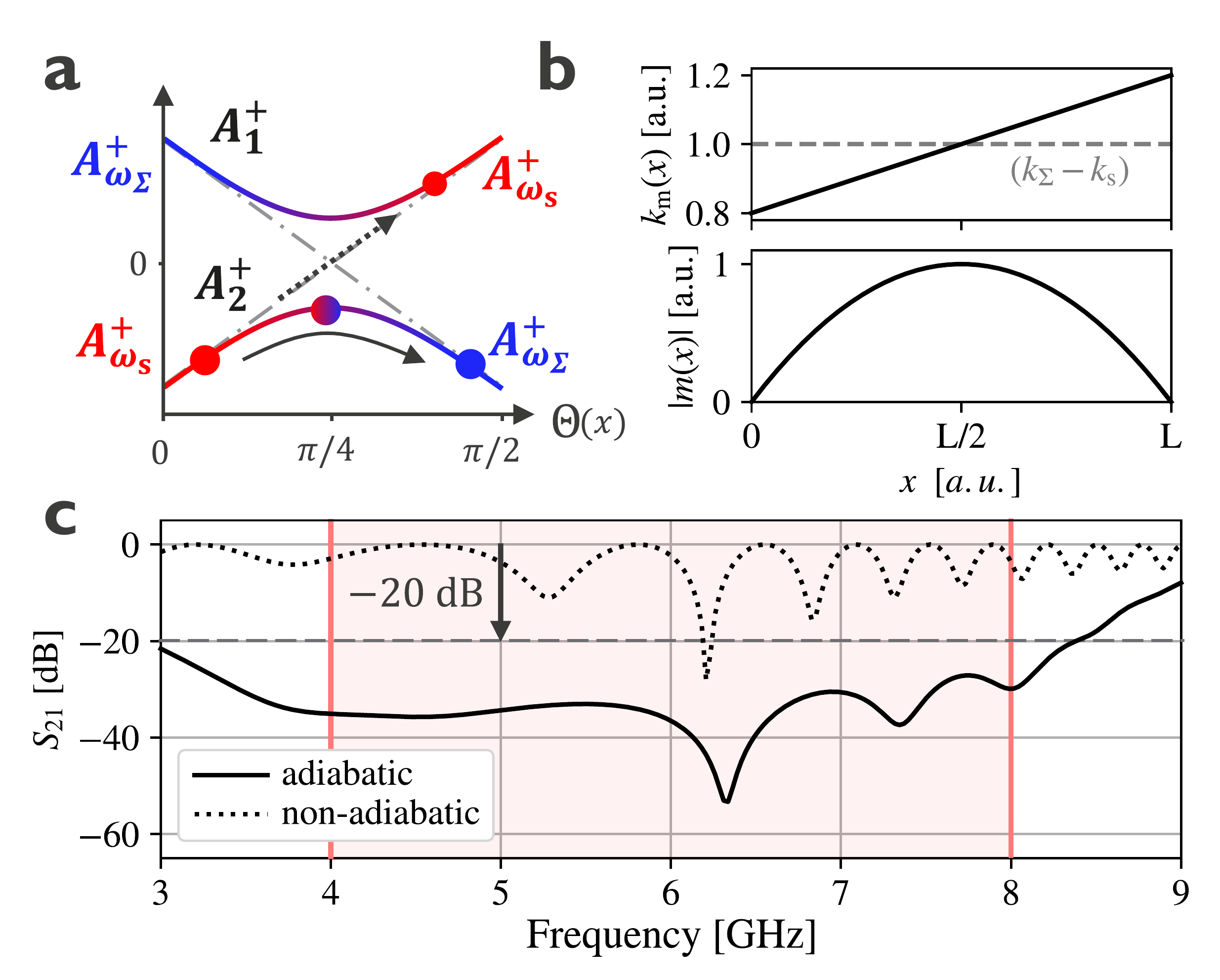}
\caption{\textbf{Adiabatic mode conversion} \textbf{(a)} The population of the lower dressed mode $A_2^+$ is converted from mode $A^{+}_{\omega_{\mathrm{s}}}$ to mode $A^{+}_{\omega_{\Sigma}}$ by adiabatically sweeping the mixing angle $\Theta$ from $0$ to $\pi/2$ (solid arrow), avoiding coupling to $A^+_{1}$ (dotted arrow).
 \textbf{(b)} Example of a quasi-adiabatic spatial variation of the modulation's amplitude $|m(x)|$ and wavenumber $k_{\mathrm{m}}(x)$, with $|m| = m_0 \left[1-\left(1-2x/L\right)^2\right]$ and $k_{m} = k_{\mathrm{m}0}\left[1-0.2\left(1-2x/L\right)\right]$. \textbf{(c)} Comparison of the isolation bandwidth obtained with Eq.~\eqref{eq:CoupledTLSRotatingFrame} for a device with and without the adiabatic modulation profile shown in (b).}
\label{fig:AdiabaticConversion}
\end{figure}

Due to the parametric coupling, the right-propagating bare modes $A_{\omega_{s}}^+$ and $A_{\omega_{\Sigma}}^+$ of the waveguide hybridize into dressed super-modes $A_{1/2}^+$, as schematized in Fig.~\ref{fig:AdiabaticConversion}a, that is
\begin{equation}
 \begin{bmatrix}
A^{+}_{1}\\ A^{+}_{2}
\end{bmatrix} 
= \begin{bmatrix} \cos \Theta
& \sin \Theta \\
- \sin \Theta & \cos \Theta
\end{bmatrix} \cdot \begin{bmatrix}
  A^{+}_{\omega_{\Sigma}}\\ \ A^{+}_{\omega_{s}}
\end{bmatrix} \hspace{5mm}
\end{equation}
with the mixing angle $\Theta(x) \in [0,\pi/2]$ defined as
\begin{equation}\label{eq:MixingAngle}
\Theta(x) = \frac{1}{2}\arctan(|g|/\Delta k^+_\Sigma)
\end{equation}
where the phase mismatch $\Delta k^+_\Sigma(x)$ and parametric coupling rate $g(x)$ are defined in Eqs.~\eqref{eq:PhaseMismatch} and \eqref{eq:TLSCouplingRate}, respectively.
In this dressed mode basis, the dynamics of the modulated waveguide translates to
\begin{equation}
\label{eq:EoMDressedFrame}
\frac{\mathrm{d}}{\mathrm{d}x} \begin{bmatrix} A^{+}_{1} \\ A^{+}_{2}
\end{bmatrix}
\approx \left\{ \frac{i\pi}{\Lambda}\begin{bmatrix} -1 & 0 \\
0 & 1 \end{bmatrix} + \frac{\mathrm{d} \Theta}{\mathrm{d}x} \begin{bmatrix} 0 & 1 \\ -1 & 0 \end{bmatrix} \right\}
\cdot 
\begin{bmatrix} A^{+}_{1} \\ A^{+}_{2} \end{bmatrix}
\end{equation}
Thus, the coupling rate between the dressed modes only depends on the rate of change of $\Theta$ over space and in the adiabatic limit where $\mathrm{d} \Theta /\mathrm{d}x = 0$, these dressed modes fully decouple.\par

In the dispersive limit, where $|\Delta k^+_\Sigma(x)|\ll|g|$, the mixing angle $\Theta \approx 0 \text{ or } \pi/2$ and the dressed modes are approximately equal to the bare modes of the unmodulated transmission line, as depicted in Fig.~\ref{fig:AdiabaticConversion}a. 
By adiabatically sweeping the mixing angle $\Theta$ from $0$ to $\pi/2$, the right-propagating signal in the bare mode at $\omega_{\mathrm{s}}$ are converted to the bare mode at $\omega_\Sigma$. To vary $\Theta$, the ratio $|g|/\Delta k^+_\Sigma$ can be varied through the amplitude $|m|$ and wavenumber $k_{\mathrm{m}}$ of the modulation, as seen in Eqs.~\eqref{eq:PhaseMismatch} and \eqref{eq:TLSCouplingRate}. For example, the spatial modulation profiles shown in Fig.~\ref{fig:AdiabaticConversion}b can be used to sweep $\Theta$ from $0$ to $\pi/2$ and produce the desired quasi-adiabatic mode conversion with $\mathrm{d}\Theta/\mathrm{d}x\ll\pi/\Lambda$.\par

The amplitude and wavenumber of the parametric modulation is thus modulated along the line, as illustrated in Fig.~\ref{fig:SystemDescription}a. As it propagates, the parametric modulation is swept through the phase-matching condition $\Delta k_\Sigma^+=0$ and adiabatically converts the right-propagating signals in the bandwidth of interest to a higher-frequency band without any back conversion, as seen in Fig.~\ref{fig:SystemDescription}d. Filtering out the higher-frequency band, we can obtain broadband isolation as a consequence of adiabatic mode conversion. Solving the system given in Eq.~\eqref{eq:CoupledTLSRotatingFrame}, the proposed scheme can achieve more than $20 \text{ dB}$ isolation over more than $4 \text{ GHz}$ bandwidth, as shown in Fig.~\ref{fig:AdiabaticConversion}c. Moreover, the adiabatic modulation is never phase-matched in the opposite direction for left-propagating signal, as indicated Fig.~\ref{fig:SystemDescription}d, so $|g|\ll\Delta k^-_\Sigma$ and $\Theta\approx 0$ all along the line. In the opposite direction, the dressed modes therefore stays dispersively equal to the bare modes and no conversion takes place, ensuring low insertion loss.

\section{Superconducting Circuit Implementation}\label{sec:Implementation}

\begin{figure}[ht]
\centering
\includegraphics[width=1.05\linewidth]{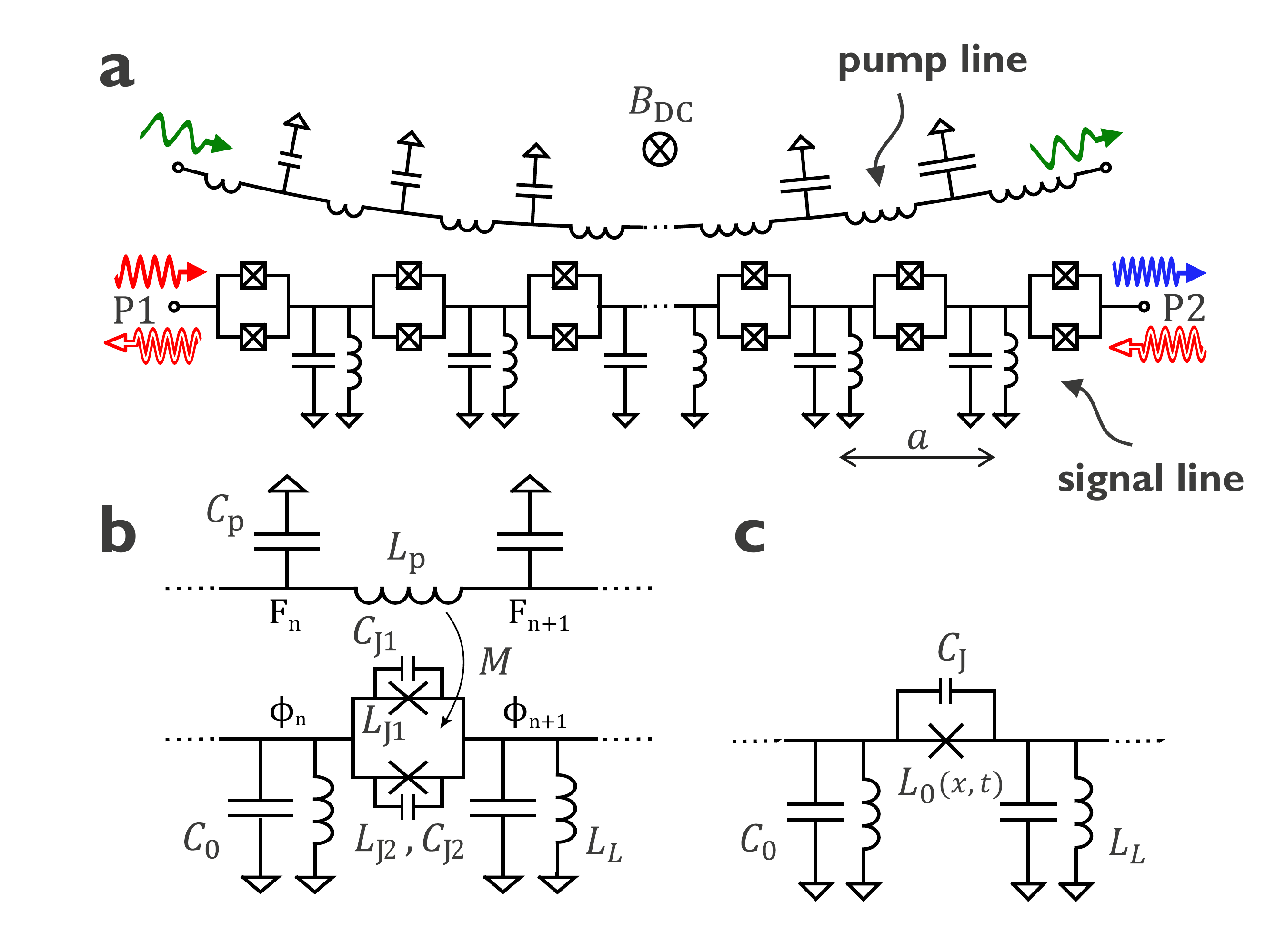}
\caption{\textbf{Circuit schematic of the proposed isolator.} \textbf{(a)} The pump propagating in the upper \textit{pump} line modulates the inductance of the lower \textit{signal} line where the signals of interest propagate. \textbf{(b)} Every unit cell in the signal line comprises a symmetric DC-SQUID, i.e. with $L_{\mathrm{J}1}=L_{\mathrm{J}2}$ and $C_{\mathrm{J}1}=C_{\mathrm{J}2}$, whose inductance is modulated by the magnetic flux threading the loop, including a contribution $\Phi_{\mathrm{DC}}$ from an external static magnetic field $B_{\mathrm{DC}}$ and a varying contribution $\Phi_{\mathrm{AC}}$ from the pump’s oscillating magnetic field. \textbf{(c)} The equivalent inductance $L_0(x,t)$ of every unit cell in the signal line hence varies over time and space as the pump propagates on the adjacent pump line.}
\label{fig:circuit}
\end{figure}

To integrate this broadband magnetless isolator using superconducting circuits, we propose the circuit implementation schematized in Fig.~\ref{fig:circuit}a, consisting of two artificial transmission lines, one supporting the propagation of the signals and another supporting a propagating pump which provides the parametric modulation described in Sec.~\ref{subsec:DirectionalParametricCoupling}. These \textit{signal} and \textit{pump} lines can be realized with lumped elements arranged in a series of $N$ subwavelength unit cells of length $a$, shown in Fig.~\ref{fig:circuit}b. The inductors and capacitors in both lines can be realized using traditional multilayer approaches \cite{Ranadive2024,Gaydamachenko2025,Macklin2015,Malnou2025,Planat2020,Roudsari2023,Pokhrel2024,Wan2021}.
Alternatively, planar capacitors can be used to minimize dielectric participation and losses \cite{Wang2025,Chang2025,Ranzani2017}. 
For compactness, the inductors $L_{\mathrm{L}}$ and $L_{\mathrm{p}}$ in both lines could also be realized with a series of Josephson junctions (JJs) operated in the linear regime, i.e. $I\ll I_{\mathrm{c}}$. Due to the nature of the conversion process, the signal and pump power are lower than in parametric amplifiers and unwanted nonlinear effects can be minimized.

Finally, superconducting quantum interference devices (SQUIDs) are placed in every unit cell of the signal line. The pump propagating in the adjacent line modulates the magnetic flux threaded by these SQUIDs resulting in a  propagating parametric modulation \cite{Zorin2019}.

\subsection{Flux-Modulated SQUIDs}\label{subsec:ModulatedSQUID}
The equivalent inductance of a DC-SQUID is a direct function of the magnetic flux $\Phi_{\mathrm{ext}}$ threading it (Appendix~\ref{appendix:SQUID}).
The magnetic flux $\Phi_{\mathrm{ext}}(x,t)$ intercepted by the SQUID comprises two terms, one constant flux $\Phi_{\mathrm{DC}}$ proportional to the external magnetic field $B_{\mathrm{DC}}$ and one varying flux $\Phi_{\mathrm{AC}}(x,t)$ proportional to the pump's power $P_{\mathrm{p}}$. The external magnetic field $B_{\mathrm{DC}}$ can be generated by an external coil or integrated on the circuit as a separate DC-flux line.
In the limit where the flux modulation induced by the pump is small, i.e. $\Phi_{\mathrm{AC}}\ll\phi_0$ where $\phi_0 = \Phi_0/2\pi$ is the reduced superconducting magnetic flux quantum, the inductance of the SQUID at location $x$ and modulated by a right-propagating pump with angular frequency $\omega_{\mathrm{p}}$ can be approximated as (Appendix~\ref{appendix:WaveEquations})
\begin{equation}\label{eq:L_0(t)}
L_{0}(x,t) = \frac{L_{\mathrm{DC}}}{1+|m|\sin\big(\omega_{\mathrm{p}}t - \int^{x}_0 k_{\mathrm{p}}(x') \ \text{d}x'\big)}
\end{equation} 
with $k_{\mathrm{p}}(x)$ the wavenumber of the pump, $L_{\mathrm{DC}} = 0.5L_{\mathrm{J}0}/ \cos\left(\Phi_{\mathrm{DC}}/2\phi_0\right)$ the unmodulated inductance and $|m|$ the modulation amplitude (Appendix~\ref{appendix:WaveEquations})
\begin{equation}\label{eq:ModulationDepth}
|m(x)| = \kappa(x) A_{\mathrm{p}} \sin\left(\frac{k_{\mathrm{p}}(x)}{2}\right) \tan\left(\frac{\Phi_{\mathrm{DC}}}{2\phi_0}\right) \ll 1
\end{equation}
where $A_{\mathrm{p}}\propto \sqrt{P_{\mathrm{p}}}$ is the dimensionless amplitude of the pump and $\kappa(x) = M(x)/L_{\mathrm{p}}(x)$ is the unit-less coupling strength between the two lines, with $L_{\mathrm{p}}(x)$ the series inductance of the pump line's unit cells and $M(x)$ the mutual inductance between the two lines (Fig.~\ref{fig:circuit}b), respectively. The flux-modulated SQUID in Fig.~\ref{fig:circuit}b hence acts as a modulated linear inductance $L_0(x,t)$ as shown in Fig.~\ref{fig:circuit}c, with a propagating modulation $m(x,t) = |m(x)|\sin\big(\omega_{\mathrm{p}}t - \int^{x}_0 k_{\mathrm{p}}(x') \ \text{d}x'\big)$. 

\subsection{Dispersion Relation}

\begin{figure}[ht]
\centering
\includegraphics[width=\linewidth]{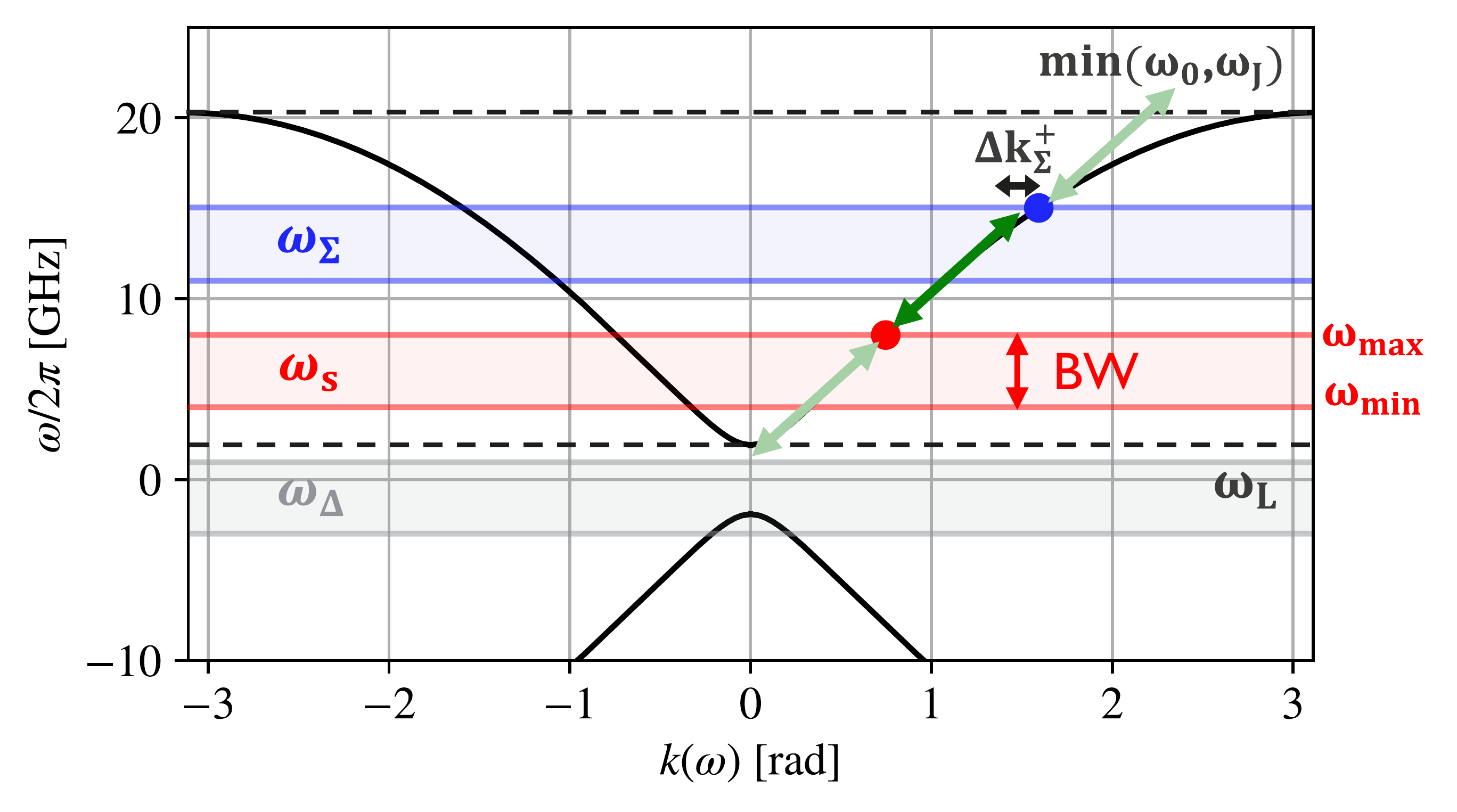}
\caption{\textbf{Dispersion engineering and isolation bandwidth.} The bandgap below $\omega_{\mathrm{L}}$ suppresses parametric coupling to lower-frequency modes for any signal within the isolation bandwidth $\text{BW}$, up to $\omega_{\mathrm{max}}<\omega_{\mathrm{L}}+\omega_p$, . At higher frequencies, the curvature of the dispersion relation, induced by the line's Bragg frequency $\omega_0$ and/or the SQUIDs' plasma frequency $\omega_{\mathrm{J}}$, creates a phase-mismatch $\Delta k_{\Sigma}^+$ between $\omega_s$ and $\omega_\Sigma$ which limits the isolation obtained for modes close to $\omega_{\mathrm{max}}$.}
\label{fig:DispersionDiagram}
\end{figure}

As discussed in Sec.~\ref{subsec:DispEng}, the dispersion of the signal line is engineered at lower frequencies, to mitigate the spurious amplification, and at higher frequencies, to form an effective two-mode system.
At lower frequencies, undesired amplification is suppressed by shunting the unit cells with an inductance $L_{\mathrm{L}}$ to ground, as shown in Fig.~\ref{fig:circuit}. These inductors create a bandgap in the dispersion curve, shown in Fig.~\ref{fig:DispersionDiagram}, which filters-out the modes below $\omega_{\mathrm{L}}=1/\sqrt{L_{\mathrm{L}} C_0}$. The inductors can be realized with geometric or kinetic inductance, or with arrays of JJs operated in their linear regime where $I\ll I_{\mathrm{c}}$. Alternatively, the shunt inductors can be replaced by series capacitors to filter out the lower frequencies. These dispersive elements can also be distributed periodically along the line rather than being placed in every single cell. \par
At higher frequencies, the nonlinear dispersion relation that prevents parametric coupling to higher frequency modes (see Sec.~\ref{subsec:DispEng}) can be implemented in two ways. One possibility is to use the Bragg frequency of the artificial transmission line. For frequencies approaching $\omega_0 = 2/\sqrt{L_{\mathrm{DC}}C_0}$, the dispersion relation flattens, as shown in Fig.~\ref{fig:DispersionDiagram}. Alternatively, this curvature can also be obtained from the plasma frequency $\omega_{\mathrm{J}}=1/\sqrt{L_{\mathrm{DC}}C_{\mathrm{J}}}$ of the SQUIDs which present a self-capacitance $C_{\mathrm{J}}= C_{\mathrm{J}1}+C_{\mathrm{J}2}$ as depicted in Fig.~\ref{fig:circuit}. \par
As seen in Fig.~\ref{fig:DispersionDiagram}, the resulting dispersion relation prevents parametric coupling except between the two sets of modes at $\{\omega_s\}$ and  $\{\omega_i\}$, hence forming a set of effective two-mode systems. Moreover, the group velocities of all the modes within the two sets are approximately equal. The adiabatic mode conversion is hence effective for every mode within the $\{\omega_s\}$ band and depletes them all simultaneously,  providing broadband microwave isolation.

\subsection{Adiabatic Flux Modulation} \label{subsec:AdiabModulation}
As explained in Sec.~\ref{subsec:AdiabModeConv}, the wavenumber and amplitude of the propagating modulation must vary gradually over space to obtain the desired quasi-adiabatic mode conversion. 
The wavenumber of the modulation can be varied by changing the wavenumber of the pump inducing the flux modulation. The wavenumber of the pump line, $k_{\mathrm{p}}(x) =  k_{\mathrm{p}0}f(x)\approx \omega_{\mathrm{p}}\sqrt{L_{\mathrm{p}}C_{\mathrm{p}}}/a$ (Appendix~\ref{appendix:WaveEquations}), is swept by increasing the product $L_{\mathrm{p}} C_{\mathrm{p}}$ from cell to cell. However, the ratio $L_{\mathrm{p}}/C_{\mathrm{p}}$ must stay fixed to maintain the same characteristic impedance $Z_{\mathrm{p}}=\sqrt{L_{\mathrm{p}}/C_{\mathrm{p}}}\approx 50\ \Omega$ along the pump line and avoid back-scattering of the pump.\par
On the other hand, the modulation amplitude $|m|$ also depends on $k_{\mathrm{p}}(x)$ but its spatial profile can be independently controlled by changing the coupling strength $\kappa(x)$ between the two lines, as seen in Eq.~\eqref{eq:ModulationDepth}. As illustrated in Fig.~\ref{fig:circuit}a, varying the distance between the two lines directly changes the mutual inductance $M(x)$ and therefore $\kappa(x)=M(x)/L_{\mathrm{p}}(x)$. The mutual inductance $M(x)$ could also be varied by changing the SQUID's area, but that approach also induces variations in $\Phi_{\mathrm{DC}}$ and $L_{\mathrm{DC}}$ leading to undesired dispersion engineering \cite{Planat2020,Roudsari2023,Gaydamachenko2025}. 

\section{Numerical Demonstration}\label{sec:Simulations}

The isolator with circuit implementation shown in Fig.~\ref{fig:circuit} was simulated in Keysight ADS, to numerically demonstrate the concepts and estimate its performance. To this end, we used a model of a flux-modulated SQUID, adapted from Ref. \cite{Shiri2024} (Appendix~\ref{appendix:ADSSQUIDModel}) to simulate the modulated transmission line in time domain and with a harmonic-balance (HB) method capturing the dynamics of the system over a wide range of frequencies. \par
The circuit parameters used in simulation are given in Tab.~\ref{tab:CircuitParameters}, together with the adiabatic modulation profiles shown in Fig.~\ref{fig:AdiabaticConversion}b. The transmission spectrum of the simulated device, obtained in ADS with HB method, is presented in Fig.~\ref{fig:TransmissionSpectrum} (colored lines) for different device lengths. For comparison, the analytical estimates obtained with Eq.~\eqref{eq:CoupledModes} are also shown in Fig.~\ref{fig:TransmissionSpectrum} (gray lines) for the same device lengths.\par

\begin{table}[!ht]
\centering
\caption{Simulated circuit parameters}
\label{tab:CircuitParameters}

\begin{tabular}{ |l|l|l| } 
\hline
$L_{\mathrm{J}1}=0.64\text{ nH}$ & $L_{\mathrm{L}}=21 \text{ nH}$  & $C_{\mathrm{p}0}=0.34\text{ pF}$\\ 
$C_{\mathrm{J}1} = 3\text{ fF}$ & $C_0=0.33 \text{ pF}$  &$L_{\mathrm{p}0}=0.87\text{ nH}$ \\ 
$L_{\mathrm{J}2} = 0.64\text{ nH}$ & $\Phi_{\mathrm{AC},0}=0.1 \Phi_{\mathrm{DC}}$ & $\omega_{\mathrm{p}}/2\pi=7\text{ GHz}$  \\
$C_{\mathrm{J}2}=3 \text{ fF}$ & $\Phi_{\mathrm{DC}}= \Phi_0/3$ & $P_{\mathrm{in}}= -100\text{ dBm}$\\
\hline
\end{tabular}
\end{table}

\subsection{Isolation \& Insertion Loss}
\begin{figure}[!ht]
\centering
\includegraphics[width=\linewidth]{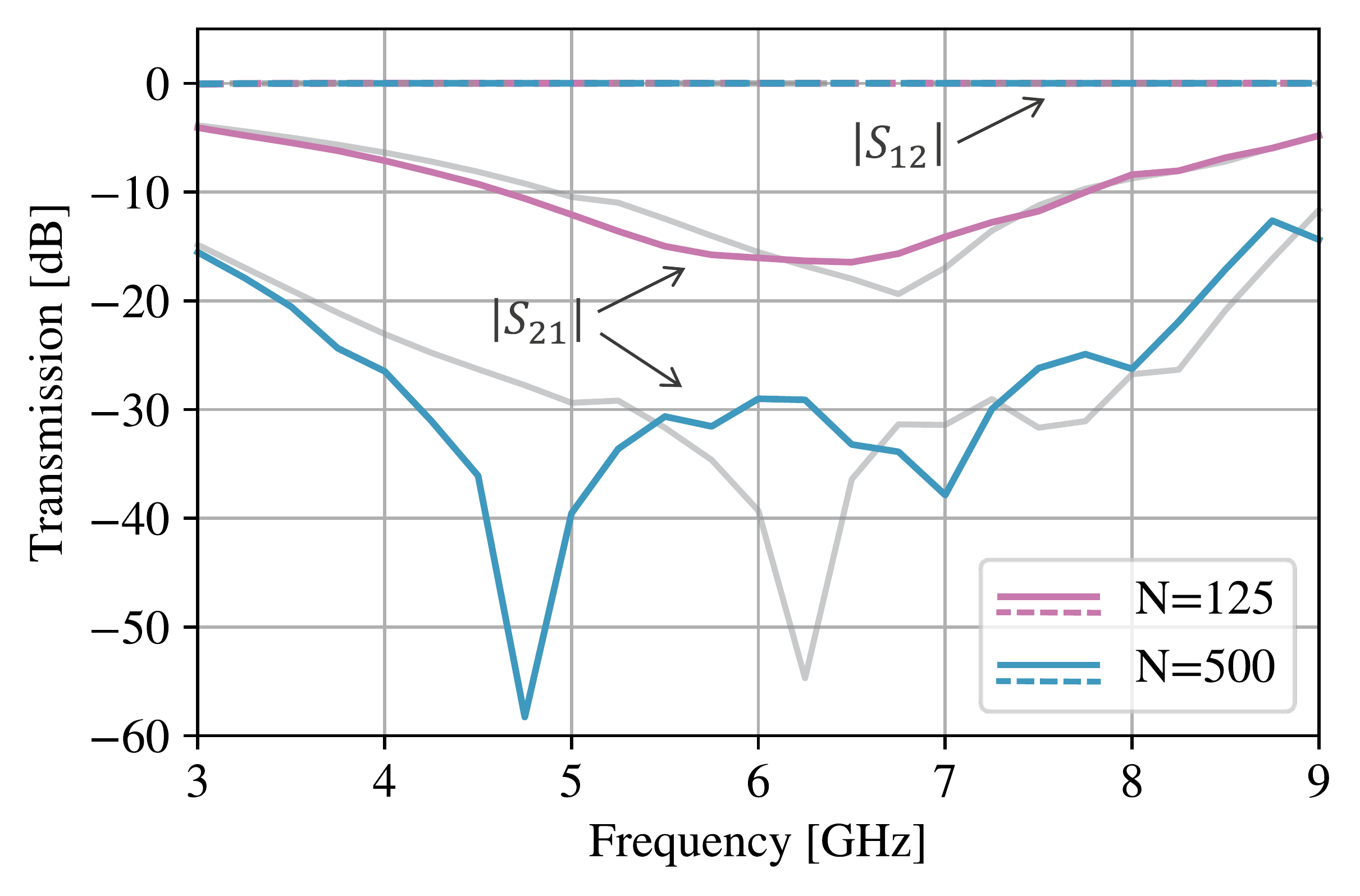}
\caption{
\textbf{Transmission spectrum} obtained by HB simulation (colored lines) and from Eq.~\eqref{eq:CoupledModes} (gray lines), for different number of unit cells $N$. In the isolating direction (solid lines), longer transmission lines provide higher isolation since the variations in modulation happen at a slower rate, which provides more adiabatic mode conversion and lower $|S_{21}|$. In the passing direction (dashed lines), the phase-mismatch is large and the parametric coupling is negligibly small regardless of the devices size leading to $|S_{12}|< 0.02 \text{ dB}$.
}
\label{fig:TransmissionSpectrum}
\end{figure}

The transmission spectrum shown in Fig.~\ref{fig:TransmissionSpectrum} shows good qualitative agreement between the analytical estimate (gray lines) obtained with Eq.~\eqref{eq:CoupledModes}, and the HB simulations (colored lines) that account for impedance mismatch, nonlinearities and higher order mixing processes. As observed in Fig.~\ref{fig:TransmissionSpectrum}, the proposed implementation achieves more than $20\text{ dB}$ isolation over a bandwidth ranging from $4$ to $8\text{ GHz}$ with $500$ unit cells, a number which is realistic in low-loss superconducting technology \cite{Macklin2015,Ranzani2017,Planat2020,Roudsari2023,Ranadive2024,Chang2025,Wang2025,Gaydamachenko2025,Malnou2025,Kow2025}.\par
Noticeably, the isolation increases with the number of cells $N$ of the device. With more cells, the spatial variation of the modulation happens at a slower rate, i.e. $\mathrm{d}\Theta/\mathrm{d}x$ in Eq.~\eqref{eq:EoMDressedFrame} decreases and the adiabaticity of the mode conversion improves, resulting in higher attenuation.
In the reverse direction of propagation, the modulation is phase-mismatched and mode conversion is limited to less than $0.02\text{ dB}$ independently of the device's length. Accounting for the back-scattering in the pump-line, simulated to be less than $-25\text{ dB}$, the mode conversion is still below $1 \text{ dB}$ over the whole isolation bandwidth.
In the reverse direction of propagation, the insertion loss is hence dominated by the dielectric losses of the chosen implementation.

\subsection{Instantaneous Bandwidth}
Two factors set an upper bound on the maximum isolation bandwidth ($\text{BW}$) achievable with this architecture, as illustrated in Fig.~\ref{fig:DispersionDiagram}. Firstly, the highest frequency $\omega_{\mathrm{max}}$ within $\text{BW}$ must be lower than $\omega_{\mathrm{max}}<\omega_{\mathrm{L}}+\omega_{\mathrm{p}}$ to avoid coupling to a lower-frequency mode which would hinder the efficiency of the adiabatic mode conversion (Sec.~\ref{subsec:AdiabModulation}). Secondly, the lowest frequency $\omega_{\mathrm{min}}$ within $\text{BW}$ must be in the pass-band of the transmission line, i.e. $\omega_{\mathrm{L}}\leq \omega_{\mathrm{min}}$, such that $\text{BW}=(\omega_{\mathrm{max}}-\omega_{\mathrm{min}})/2\pi<\omega_{\mathrm{p}}/2\pi$.
The isolation bandwidth is then maximized by increasing the pump frequency $\omega_{\mathrm{p}}$, while the bandgap below $\omega_{\mathrm{L}}$ ensures that undesired parametric amplification of the modes at $\{\omega_{\mathrm{s}}\}<\omega_{\mathrm{p}}$ is suppressed (Sec.~\ref{subsec:DispEng}). \par

Besides this upper bound, the isolation bandwidth is limited by the curvature of the dispersion relations for frequencies close to the lower cut-off frequency $\omega_{\mathrm{L}}$ and the higher Bragg frequency $\omega_0=2/\sqrt{L_{\mathrm{DC}} C_0}$ as shown in Fig.~\ref{fig:DispersionDiagram}. The nonlinear dispersion relation in this range of frequencies disturbs the phase-matching, effectively reducing the efficiency of the adiabatic mode conversion that provides the isolation.
As seen in Fig.~\ref{fig:TransmissionSpectrum}, the simulated circuit provides more than $20 \text{ dB}$ isolation for signals between $4$ and $8\text{ GHz}$, hence achieving a $4 \text{ GHz}$ instantaneous isolation bandwidth when using a pump at $7 \text{ GHz}$.
This $\text{BW}$ could be extended with a larger $\omega_{\mathrm{p}}$ and by raising the Bragg frequency $\omega_0$ at the cost of increasing the number of cells $N$ to maintain the same electrical length. \par

\subsection{Line Inhomogeneities}
\begin{figure}[]
\centering
\includegraphics[width=
\linewidth]{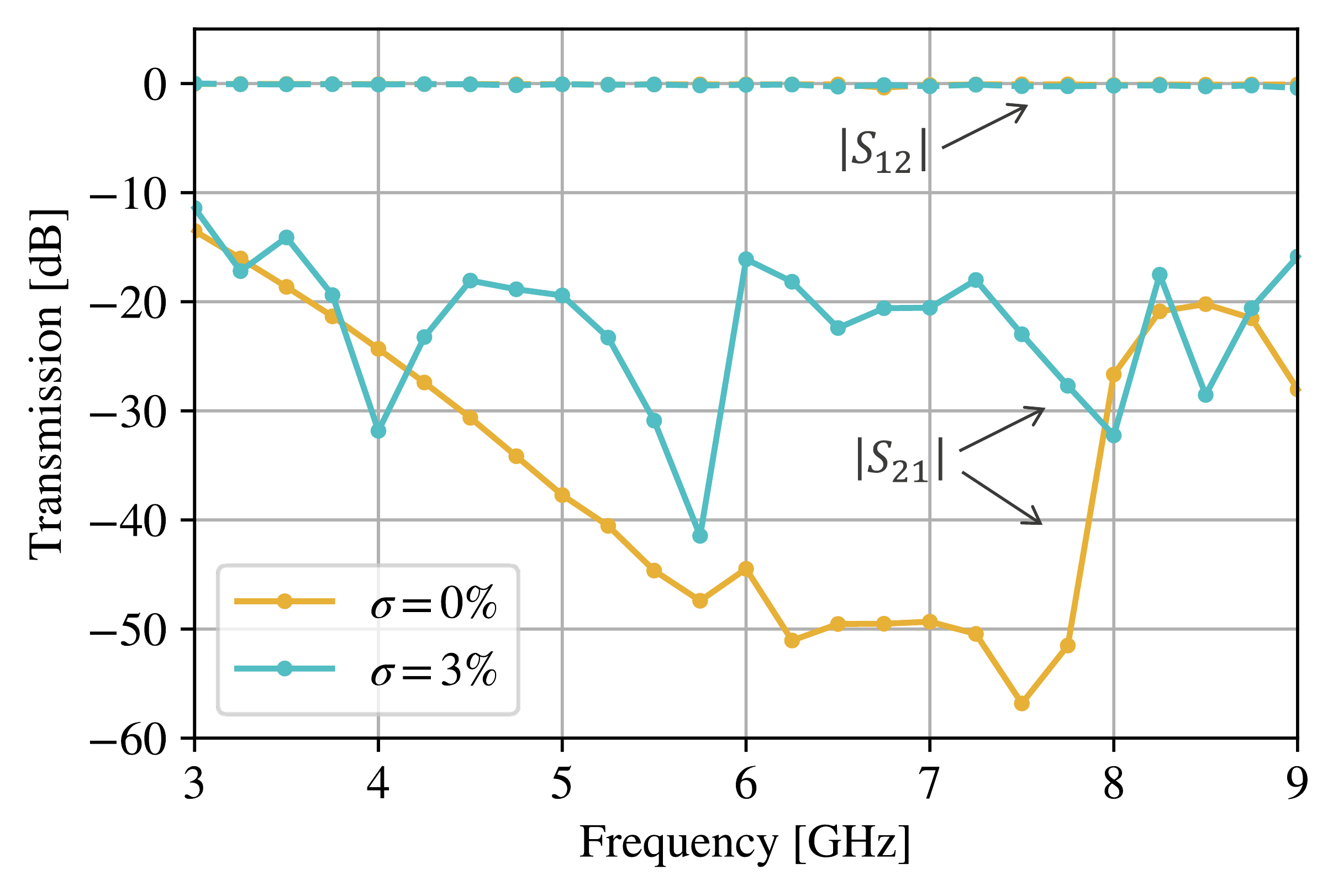}
\caption{\textbf{Transmission spectrum} of the proposed isolator obtained by time-domain simulation with the seven first parameters in Tab.~\ref{tab:CircuitParameters} varying from cell to cell following a normal distribution with standard deviation $\sigma=0\%$ (yellow lines) and $\sigma= 3\%$ (green lines).}
\label{fig:Inhomogeneities}
\end{figure}

Finally, the impact of line inhomogeneity, induced by device-fabrication methods, was also studied in time-domain simulations to verify the robustness of our approach. To that end, the isolator was simulated with the seven first circuit parameters from Table~\ref{tab:CircuitParameters} varying from cell-to-cell following a normal distribution with a standard deviation of $\sigma=3\%$, which is a realistic spread in current state-of-the-art fabrication \cite{Wan2021,Kissling2023,VanDamme2024}. The impact of these variations and resulting asymmetries was most important for modes close to the Bragg frequency $\omega_0$ where impedance variations are largest and back-scattering limits the achievable isolation. Nonetheless, this effect can be attenuated by raising $\omega_0$ and increasing the number of unit cells to maintain a certain electrical length. Raising $\omega_0/2\pi$ from $20\text{ GHz}$ to $40 \text{ GHz}$, $N$ from $500$ to $750$ and reducing $\omega_{\mathrm{p}}/2\pi$ from $7$ to $6 \text{ GHz}$, the circuit including parameter variability with $\sigma=3\%$ achieved more than $16 \text{ dB}$ isolation and less $0.4\text{ dB}$ insertion loss for any signal from $4$ to $8 \text{ GHz}$, as shown in Fig.~\ref{fig:Inhomogeneities}. These results indicate that the proposed circuit implementation could support parameters variation due to fabrication while achieving state-of-the-art performances.\par

\section{Conclusion}\label{sec:Discussion}
In this work, we propose a novel architecture
to produce broadband magnetless isolation and potentially replace ferrite-based isolators in superconducting circuits. Leveraging parametric modulations and dispersion engineering, we realize an effective two-mode system with direction-dependent coupling and matched group velocities inside a single transmission line. This configuration enables adiabatic mode conversion over a wide instantaneous bandwidth, providing broadband isolation.
Using Lagrangian analysis and circuit simulations, we demonstrated isolation exceeding $20\text{ dB}$ over $4-8\text{ GHz}$, which is on par with the state-of-the-art and could be further optimized by adjusting the circuit parameters and modulation profiles.\par
The proposed circuit implementation offers several key advantages for on-chip integration. 
The $50\ \Omega$-matched artificial transmission line offers direct access to the effective two-mode system without any splitters, combiners, or other matching networks. 
The separate pump line eliminates the need for directional couplers and prevents the pump from reaching the device under test. 
The moderate number of unit cells in the proposed implementation results in a reduced footprint and higher device yield. 
Finally, our variability analysis confirms that the adiabatic sweep through the phase-matching condition relaxes the precision required in fabrication and parameter targeting. 
With these features, the proposed broadband magnetless isolator presents a scalable alternative to ferrite isolators that could be co-integrated on-chip and support the development of large-scale superconducting quantum processors and other ultra-low-noise cryogenic systems.

\begin{acknowledgments}
This work was supported in part by the imec Industrial Affiliation Program on Quantum Computing. M.D. acknowledges the support of the Research Foundation - Flanders through the Strategic Basic Research PhD program (grant no. 1SHHM24N). We thank K. Moors and A. Potočnik for their insightful comments on this work.
\end{acknowledgments}

\bibliography{references}

\appendix

\section{Wave Equations}\label{appendix:WaveEquations}

Consider two lossless transmission lines and their equivalent circuit given in Fig.~\ref{fig:circuit}. The \textit{signal line}, i.e. the transmission line where the signals of interest propagate, is composed of a series of $N$ identical unit cells of size $a$ (Fig.~\ref{fig:circuit}a). Each cell comprises a DC-SQUID formed with two JJs in parallel, a shunt capacitor $C_0$ and a shunt inductor $L_{\mathrm{L}}$ to ground. The second adjacent \textit{pump line} supports the propagation of an electromagnetic wave called the \textit{pump} whose magnetic field modulates the magnetic flux threading the SQUIDs in the first signal line. Similarly, it is made of a series of unit cells composed of a series inductance $L_{\mathrm{p}}(x)$, a shunt capacitor $C_{\mathrm{p}}(x)$ and some inductive coupling $M(x)$ to the SQUID in the signal line. The values of theses lumped elements in the pump line generally vary from cell to cell, as indicated by their dependency on the dimensionless spatial variable $x$ normalized to the size $a$ of a unit cell.\par

The dynamics of the circuit can be described in terms of the independent node fluxes $F_n$ and $\phi_n$ (see Fig.~\ref{fig:circuit}b) defined as
\begin{subequations}
  \begin{equation}
  F_n(t) \equiv \int_{-\infty}^{t} v_{\mathrm{p},n}(\tau) \ \mathrm{d}\tau
  \end{equation}
  \begin{equation}
  \phi_n(t) \equiv \int_{-\infty}^{t} v_{\mathrm{s},n}(\tau) \ \mathrm{d}\tau
  \end{equation} 
\end{subequations}
with the node voltages $v_{\mathrm{p},n}$ and $v_{\mathrm{s},n}$ of $n$-th cell of the pump and signal lines, respectively.

With these generalized coordinates, the Lagrangian of the circuit consist of two contributions, from the pump and signal lines
\begin{equation}
  \mathcal{L} = \mathcal{L}_{\mathrm{p}} + \mathcal{L}_{\mathrm{s}} 
\end{equation}
The Lagrangian of the pump line is found by subtracting the potential from the kinetic energy for every unit cell, shown in Fig.~\ref{fig:circuit}b. 
The kinetic energy of capacitors and potential energy of the linear inductors are found as \cite{Vool2017}
\begin{equation}
K_C = \frac{C}{2}\Delta\dot{\phi}^2 \hspace{5mm} \mathrm{and} \hspace{5mm} U_L = \frac{1}{2L}\Delta\phi^2 
\end{equation}
where $\Delta\phi$ is the phase difference across the inductive or capacitive element. Therefore we find the Lagrangian of the pump line as
\begin{equation}
  \mathcal{L}_{\mathrm{p}} 
  = \sum_{n=0}^N\Bigg[ \frac{C_{\mathrm{p}}(n) }{2}(\dot{F}_n)^2 - \frac{\left(F_{n+1}-F_n\right)^2}{2L_{\mathrm{p}}(n)}\Bigg]
\end{equation}
with the boundary conditions $F_1=F_{\mathrm{in}}$ and $F_N=F_{\mathrm{out}}$.
For the signal line, the potential energy of a symmetric DC-SQUID with $L_{\mathrm{J}1}=L_{\mathrm{J}2}=L_{\mathrm{J}0}$ is found as (Appendix \ref{appendix:SQUID})
\begin{equation}
  U_{\mathrm{SQUID}} = -2E_{\mathrm{J}0} \cos{\left(\frac{\Phi_{\mathrm{ext}}}{2\phi_0}\right)} \cos\left(\frac{\Delta\phi}{\phi_0}\right)
\end{equation}
with the Josephson energy defined as $E_{\mathrm{J}0} =\phi_0^2/L_{\mathrm{J}0}$. The DC-SQUID hence acts as a single Josephson junction whose critical current varies as a function of the external magnetic flux $\Phi_{\mathrm{ext}}$ threading the SQUID. The kinetic energy of a symmetric DC-SQUID arises from the charging energy of the two junctions' capacitance $C_{\mathrm{J}1}=\ C_{\mathrm{J}2}=C_{\mathrm{J}}/2$ (Fig.~\ref{fig:circuit}b,c)
\begin{equation}
  K_{\mathrm{SQUID}} = \frac{C_{\mathrm{J}}}{2} \Delta\dot{\phi}^2
\end{equation}

From these equations, the Lagrangian of the signal line follows as 
\begin{align}
\mathcal{L}_{\mathrm{s}} = & \sum_{n=0}^N \Bigg[ \frac{C_0 }{2}(\dot{\phi}_n)^2 + \frac{C_{\mathrm{J}} }{2}(\dot{\phi}_{n+1}-\dot{\phi}_{n})^2 - \frac{1}{2L_{\mathrm{L}}}\phi_n^2 \\
& + 2E_{\mathrm{J}0} \cos\left(\frac{\Phi_{\mathrm{ext}}(n,t)}{2\phi_0}\right) \cos\left(\frac{\phi_{n+1}-\phi_{n}}{\phi_0}\right) \Bigg] \nonumber
\end{align}
with the boundary conditions $\phi_0=\phi_{\mathrm{in}}$, $\phi_N=\phi_{\mathrm{out}}$ and
where the coupling between the two lines arises through the external magnetic flux $\Phi_{\mathrm{ext}}$ intercepted by the SQUIDs. The magnetic flux $\Phi_{\mathrm{ext}}$ has two components, that is
\begin{equation}\label{eq:Phi_ext,n_Appendix}
  \Phi_{\mathrm{ext}}(x,t) = \Phi_{\mathrm{DC}} + \Phi_{\mathrm{AC}}(x,t)
\end{equation}
where $\Phi_{\mathrm{DC}}$ is a constant flux resulting from the external magnetic field $B_{\mathrm{DC}}$, and $\Phi_{\mathrm{AC}}(x,t)$ a oscillating flux resulting from the pump propagating on the adjacent line.
The varying flux $\Phi_{\mathrm{AC}}$ is therefore induced by the current flowing in the pump line through the inductance $L_{\mathrm{p}}$, found as 
\begin{equation}
I_{\mathrm{p},n}(t) = -\left( F_{n+1}-F_n\right)/L_{\mathrm{p}}(n)
\end{equation}
yielding the time-varying flux 
\begin{equation}\label{eq:PhiACn_Appendix}
\Phi_{\mathrm{AC}}(n,t) = M(n) I_{\mathrm{p},n}(t) = -\kappa(n)\left( F_{n+1}-F_n\right)
\end{equation}
with $M(x)$ being the mutual inductance between the lines as depicted in Fig.~\ref{fig:circuit}b and $\kappa(x) = M(x)/L_{\mathrm{p}}(x)$ the unit-less coupling strength between the two lines.\par

\subsection{Pump Transmission Line}
In the case of weak coupling between the pump and signal lines, i.e. $|\kappa(x)|^2 \ll 1$ and the pump is much larger in amplitude than the signal and the idler \footnote{For signals in the $4$ to $8 \text{ GHz}$ band, the Manley-Rowe relations \cite{Manley1956} guarantee that the idler generated during the mode conversion process can be at most $5 \text{ dB}$ larger than the signal, i.e. $G_{\mathrm{s}\rightarrow \mathrm{i}} = f_{\mathrm{i}}/f_{\mathrm{s}} = 1 + f_{\mathrm{s}}/f_{\mathrm{p}} <3$.}, the back-action of the signal line on the pump line can be neglected. The pump line hence decouples from the signal line and the Euler-Lagrange equation reads
\begin{equation}
\frac{\mathrm{d}}{\mathrm{d} t}\left( \frac{\partial \mathcal{L}}{\partial \dot{F}_n}\right) - \frac{\partial \mathcal{L}}{\partial F_n} \approx \frac{\mathrm{d}}{\mathrm{d} t}\left( \frac{\partial \mathcal{L}_{\mathrm{p}}}{\partial \dot{F}_n}\right) - \frac{\partial \mathcal{L}_{\mathrm{p}}}{\partial F_n} = 0
\end{equation}
which gives
\begin{align}
  C_{\mathrm{p}}(n)\ddot{F}_n +&\frac{1}{L_{\mathrm{p}}(n-1)}\left(F_{n}-F_{n-1}\right) \\
  &- \frac{1}{L_{\mathrm{p}}(n)}\left(F_{n+1}-F_n\right) = 0 \nonumber
\end{align}
with $n =1,2, \dots ,N-1$. Since the variation of $C_{\mathrm{p}}(x)L_{\mathrm{p}}(x)$ is adiabatic, the inductance variation between adjacent cells is assumed to be negligibly small, i.e. $L_{\mathrm{p}}(n-1)\approx L_{\mathrm{p}}(n)$, and the Euler-Lagrange equation reads
\begin{equation}\label{eq:PumpDiscreteEoM_Appendix}
  \ddot{F}_n - \nu_{\mathrm{p}}^2 \left(F_{n+1} - 2 F_n + F_{n-1} \right) = 0
\end{equation}
with the phase velocity $\nu_{\mathrm{p}}(x) \equiv 1/\sqrt{L_{\mathrm{p}}(x) C_{\mathrm{p}}(x)}$, which is normalized to the length $a$ of one unit cell and that varies slowly with position $x$.\par

From the Bloch-Floquet theorem, we know that periodic structure supports the propagation of plane waves. Similarly, the solution to the wave equation propagating in the quasi-periodic pump line can be expressed as a superposition of right- and left-propagating plane waves 
\begin{align}\label{eq:AnsatzF_Appendix}
F_n(t) = i\sqrt{\frac{\hbar}{4\pi}} \int_{-\infty}^{\infty} \frac{\sqrt{Z_0(\omega_{\mathrm{p}})}}{\sqrt{|\omega_{\mathrm{p}}|}}\Big\{ B^{+}_{\omega_{\mathrm{p}}} & e^{-i\left[\omega_{\mathrm{p}} t - \theta_{\mathrm{p}} \right]} \\
+ B^{-}_{\omega_{\mathrm{p}}}& e^{-i\left[ \omega_{\mathrm{p}} t + \theta_{\mathrm{p}} \right]} \Big\} \mathrm{d}\omega_{\mathrm{p}} \nonumber
\end{align}
where $Z_0(\omega)$ is the characteristic impedance of the pump line, $\theta_{\mathrm{p}}(\omega_{\mathrm{p}},n) = \sum_{x=0}^{n} k_{\mathrm{p}}(\omega_{\mathrm{p}},x)$ is the accumulated phase of the pump at the $n$-th cell and $k_{\mathrm{p}}(\omega_{\mathrm{p}}, x)$ is the pump's wavenumber, normalized to the length $a$ of a unit cell. In this decomposition, the complex envelopes are normalized to be proportional to the number of photons in the mode, i.e. $|B^{\pm}_{\omega_{\mathrm{p}}}|^2\propto \bar{n}_{\omega_{\mathrm{p}}}^\pm $. Moreover, the complex envelopes are conjugate symmetric since $F_n$ is a real function. i.e. $B^{\pm}_{-\omega_{\mathrm{p}}} = \left(B^{\pm}_{\omega_{\mathrm{p}}}\right)^*$.\par

Substituting Eq.~\eqref{eq:AnsatzF_Appendix} in Eq.~\eqref{eq:PumpDiscreteEoM_Appendix}, the right- and left-propagating waves satisfy the following dispersion relation
\begin{equation}
  \left[\omega_{\mathrm{p}}^2 + \nu_{\mathrm{p}}^2 (e^{ik_{\mathrm{p}}} - 2 + e^{-i k_{\mathrm{p}}} )\right] = 0 
  \end{equation}
where the phase variation of the pump across a unit cell is approximated as $\theta_{\mathrm{p}}(n)-\theta_{\mathrm{p}}(n-1)\approx\theta_{\mathrm{p}}(n+1)-\theta_{\mathrm{p}}(n) \approx k_{\mathrm{p}}(\omega_{\mathrm{p}},n)$.
The normalized wave-number of the pump line is therefore found as 
\begin{equation}\label{eq:PumpWavenumber_Appendix}
  k_{\mathrm{p}}(\omega_{\mathrm{p}},x) = 2\arcsin\left( \frac{\omega_{\mathrm{p}}}{2\nu_{\mathrm{p}}(x)}\right) 
\end{equation}
where the spatial dependency enters through the normalized phase-velocity $\nu_{\mathrm{p}}(x)$ of the pump line.

\subsection{Signal Transmission Line}
As discussed in the previous section, the dynamics of the pump and signal lines decouple in the limit where the coupling between the two lines $|\kappa(x)|^2$ is weak and the pump is large compared to the signal and idler. The Euler-Lagrange equation then reads
\begin{equation}
  \frac{\mathrm{d}}{\mathrm{d} t}\left( \frac{\partial \mathcal{L}}{\partial \dot{\phi}_n}\right) - \frac{\partial \mathcal{L}}{\partial \phi_n} \approx \frac{\mathrm{d}}{\mathrm{d} t}\left( \frac{\partial \mathcal{L}_{\mathrm{s}}}{\partial \dot{\phi}_n}\right) - \frac{\partial \mathcal{L}_{\mathrm{s}}}{\partial \phi_n} = 0
\end{equation}
which gives
\begin{align}
  &0 = \ C_0 \ddot{\phi}_n - C_{\mathrm{J}}\left( \ddot{\phi}_{n+1} - 2\ddot{\phi}_{n} + \ddot{\phi}_{n-1}\right) + \frac{1}{L_{\mathrm{L}}} \phi_n \label{eq:ELSignal_Appendix} \\
  & + I_{\mathrm{c},n-1}(t) \sin\left(\frac{\phi_n - \phi_{n-1}}{\phi_0} \right) - I_{\mathrm{c},n}(t) \sin\left(\frac{\phi_{n+1} - \phi_{n}}{\phi_0}\right) \nonumber
\end{align}
with $n =1,2,\dots ,N-1$ and the boundary conditions $\phi_0=\phi_{\mathrm{in}}$ and $\phi_N=\phi_{\mathrm{out}}$. The time-dependent critical currents of the SQUIDs are given as
\begin{equation}\label{eq:SQUIDCriticalCurrent_Appendix}
  I_{\mathrm{c},x}(t) = 2 I_{\mathrm{c}0}\cos\left(\frac{\Phi_{\mathrm{ext}}(x,t)}{2\phi_0}\right)
\end{equation}
with $I_{\mathrm{c}0}=E_{\mathrm{J}0}/\phi_0$, the critical currents of the two junctions forming the SQUID and $\Phi_{\mathrm{ext}}(n,t)$ as given in Eq.~\eqref{eq:Phi_ext,n_Appendix}. 
In the limit where the phase across the SQUID is much smaller than a flux quantum, i.e. $\phi_{n+1}-\phi_n\ll\phi_0 $ or equivalently $I/I_{\mathrm{c}0} \ll 1$, the Euler-Lagrange equations in Eq.~\eqref{eq:ELSignal_Appendix} can be linearized as 
\begin{align}
  \label{eq:DiscreteSignalEqs_Appendix}
  0 =\ &C_0 \ddot{\phi}_n - C_{\mathrm{J}}\left( \ddot{\phi}_{n+1} -2 \ddot{\phi}_n + \ddot{\phi}_{n-1} \right) + \frac{1}{L_{\mathrm{L}}} \phi_n \\
  &+ \frac{1}{L_{0}(n-1,t)} \left(\phi_n - \phi_{n-1}\right) - \frac{1}{L_{0}(n,t)} \left(\phi_{n+1} - \phi_{n} \right) \nonumber
\end{align}
with the time varying SQUID's inductance $L^{-1}_{0}(x,t)= I_{\mathrm{c},x}(t)/\phi_0$, or substituting Eq.~\eqref{eq:SQUIDCriticalCurrent_Appendix}
\begin{equation}\label{eq:ModulatedInductanceSQUID_Appendix}
L_{0}(x,t) = \frac{0.5 L_{\mathrm{J}0}}{\cos(\Phi_{\mathrm{ext}}(x,t)/2\phi_0)}
\end{equation}
where $L_{\mathrm{J}0}$ is the inductance of the individual junctions forming the SQUID.

\subsection{Flux-Modulated SQUIDs}
In the limit where the flux-modulation induced by the pump is small, i.e. $\Phi_{\mathrm{AC}}\ll\Phi_0$, the expression from Eq.~\eqref{eq:ModulatedInductanceSQUID_Appendix} can be approximated as
\begin{equation}\label{eq:WeaklyModulatedInductanceSQUID_Appendix}
  L_{0}(x,t) \approx \frac{L_{\mathrm{DC}}}{1-\tan\left(\frac{\Phi_{\mathrm{DC}}}{2\phi_0}\right) \frac{\Phi_{\mathrm{AC}}(x,t)}{2\phi_0}}
\end{equation}
with the unmodulated inductance
\begin{equation}
  L_{\mathrm{DC}}=\frac{0.5 L_{\mathrm{J}0}}{ \cos\left(\Phi_{\mathrm{DC}}/2\phi_0\right)}
\end{equation}
Considering a right-propagating pump with angular frequency $\omega_{\mathrm{p}}$, power $P_{\mathrm{p}}\gg P_{in}$ and dimensionless amplitude $A_{\mathrm{p}}\propto \sqrt{P_{\mathrm{p}}}$, the node fluxes are found from Eq.~\eqref{eq:AnsatzF_Appendix}, as 
\begin{equation}
  F_n = \phi_0 A_{\mathrm{p}} \cos(\omega_{\mathrm{p}}t-\theta_{\mathrm{p}}) 
\end{equation}
with the accumulated phase $\theta_{\mathrm{p}}(n) = \sum_{x=0}^{n} k_{\mathrm{p}}(x) $ and the normalized pump's wavenumber $k_{\mathrm{p}}$ given in Eq.~\eqref{eq:PumpWavenumber_Appendix}. Due to the negligible back-action on the pump, $A_{\mathrm{p}}$ can be taken as a constant. Substituting in Eq.~\eqref{eq:PhiACn_Appendix}, the induced varying flux is found as
\begin{equation}
  \Phi_{\mathrm{AC}}(n,t) 
  = -2\kappa (n)\phi_0 A_{\mathrm{p}} \sin\Big(\frac{k_{\mathrm{p}}(n)}{2}\Big) \sin\big(\omega_{\mathrm{p}}t - \bar{\theta}_{\mathrm{p}}(n)\big)
\end{equation}
where we introduce $\bar{\theta}_{\mathrm{p}}(n) = \left[ \theta_{\mathrm{p}}(n+1)+\theta_{\mathrm{p}}(n)\right]/2$.
Assuming a coupling strength $\kappa\leq 0.2$, the pump power required to generate the desired flux-modulation given in Table \ref{tab:CircuitParameters} is hence around $-80 \text{dBm}$.\par 
Substituting for $\Phi_{\mathrm{AC}}$ in Eq.~\eqref{eq:WeaklyModulatedInductanceSQUID_Appendix}, the inductance of the SQUID with small flux-modulation is found as
\begin{equation}\label{eq:ModulatedInductance_Appendix}
L_{0}(n,t) = \frac{L_{\mathrm{DC}}}{1+|m(n)|\sin\big(\omega_{\mathrm{p}}t - \bar{\theta}_{\mathrm{p}}(n)\big)}
\end{equation} 
with the modulation depth $|m|$ given as
\begin{equation}
|m(n)| = \kappa(n) A_{\mathrm{p}} \sin\left(\frac{k_{\mathrm{p}}(n)}{2}\right) \tan\left(\frac{\Phi_{\mathrm{DC}}}{2\phi_0}\right) \ll 1
\end{equation}
As shown in Fig.~\ref{fig:circuit}c, the flux-modulated DC-SQUID from Fig.~\ref{fig:circuit}b is hence approximated as a time-varying linear inductance $L_0(x,t) = L_{\mathrm{DC}} \ [1+m(x,t)]^{-1}$ given in Eq.~\eqref{eq:ModulatedInductance_Appendix}.

\subsection{Parametric Coupling}
For an inductance with a propagating modulation $m(n,t)= |m(n)|\sin(\omega_{\mathrm{p}} t- \bar{\theta}_{\mathrm{p}}(n))$, see Eq.~\eqref{eq:ModulatedInductance_Appendix}, the dynamics of the transmission line given in Eq.~\eqref{eq:DiscreteSignalEqs_Appendix} become
\begin{align} \label{eq:EoM_m(x,t)_Appendix}
   \frac{1}{\nu^2 } \ddot{\phi}_n - \frac{1}{\omega_{\mathrm{J}}^2}\left( \ddot{\phi}_{n+1} -2 \ddot{\phi}_n + \ddot{\phi}_{n-1} \right) + \frac{\omega_{\mathrm{L}}^2}{\nu^2} \phi_n & \\
  - \left(\phi_{n+1}-2\phi_n + \phi_{n-1}\right) =  m(n,t) &\left(\phi_{n+1} - \phi_{n} \right) \nonumber \\
   \hspace{43mm} - m(n-1,t) & \left(\phi_{n} - \phi_{n-1} \right) \nonumber
\end{align}
where we introduced the phase velocity $\nu=1/\sqrt{L_{\mathrm{DC}}C_0}$ normalized to the length $a$ of the cell, the SQUIDs' plasma frequencies $\omega_{\mathrm{J}}=1/\sqrt{L_{\mathrm{DC}}C_{\mathrm{J}}}$ and the lower cut-off frequency $\omega_{\mathrm{L}}=1/\sqrt{L_{\mathrm{L}}C_0}$.

\subsubsection*{Unmodulated Transmission Line}
For the unmodulated line, i.e. when no pump is applied and $|m|=0$, the equations of motion simplify to
\begin{align} \label{eq:UnModulatedEoM_Appendix}
  \frac{1}{\nu^2 } & \ddot{\phi}_n - \frac{1}{\omega_{\mathrm{J}}^2}\left( \ddot{\phi}_{n+1} -2 \ddot{\phi}_n + \ddot{\phi}_{n-1} \right) + \frac{\omega_{\mathrm{L}}^2}{\nu^2} \phi_n \\
   & \hspace{20mm} - \left(\phi_{n+1}-2\phi_n + \phi_{n-1}\right)= 0 \nonumber
\end{align}
The solutions to these wave equations are again a superposition of right- and left-propagating waves,
\begin{align} \label{eq:AnsatzPhi_Appendix}
\phi_{n,0}(t) = i\sqrt{\frac{\hbar}{4\pi}} \int_{-\infty}^{\infty} \mathrm{d}\omega\ \frac{\sqrt{Z_0(\omega)}}{\sqrt{|\omega|}}\Big\{ A^{+}_{\omega} e^{-i\left[\omega t - \theta \right]}& \\
 + A^{-}_{\omega} e^{-i\left[ \omega t + \theta \right]} \Big\} \nonumber&
\end{align}
with the signal line's characteristic impedance $Z_0(\omega)$, the accumulated phase $\theta(\omega,n) = \sum_{x=0}^{n} k(\omega,x)$, the normalized wavenumber $k(\omega,x)$ and the conjugate symmetric Fourier coefficients $A^{\pm}_{-\omega} = \left(A^{\pm}_{\omega}\right)^*$ that are proportional to the number of photon in the mode, i.e. $|A^{\pm}_{\omega}|^2\propto \bar{n}_{\omega}^\pm$.\par 
Considering that the unit cells in the signal line are all identical, the normalized wavenumber $k(\omega)$ is independent of $x$, so that $\theta(\omega,n)=n k(\omega)$. In this case, substituting the ansatz of Eq.~\eqref{eq:AnsatzPhi_Appendix} in Eq.~\eqref{eq:UnModulatedEoM_Appendix}, the dispersion relation of the signal line reads
\begin{equation}
  \left[\frac{\omega^2}{\nu^2}\Big(1-\frac{\omega_{\mathrm{L}}^2}{\omega^2} \Big) + \Big(1 -\frac{\omega^2}{\omega_{\mathrm{J}}^2}\Big) (e^{ik} - 2 + e^{-i k} ) \right]= 0 
\end{equation}
and the normalized wave-number of the signal line is found as 
\begin{equation}\label{eq:SignalWavenumber_Appendix}
  k(\omega) = 2\arcsin\left( \frac{\omega}{2\nu} \sqrt{\frac{1-\omega_{\mathrm{L}}^2/\omega^2}{1-\omega^2/\omega_{\mathrm{J}}^2}}\right) 
\end{equation}
where $2\nu=\omega_0$ is the Bragg frequency of the periodic artificial transmission line.
\subsubsection*{Modulated Transmission Line}
When applying a pump, i.e. for $|m|\neq0$, the right and left-propagating modes are no longer the normal modes of the modulated transmission line. Parametric modulations hence creates coupling between the right and left-propagating waves. When expanding the node fluxes in the normal mode basis of the unmodulated system, the amplitude of these left and right-propagating modes are now function of position due to the modulation-induced coupling, that is 
\begin{align} \label{eq:BareModeExpansion_Appendix}
& \phi_n(t) = i\sqrt{\frac{\hbar}{4\pi}} \int_{-\infty}^{\infty} \mathrm{d}\omega\ \frac{\sqrt{Z_0}}{\sqrt{|\omega|}}\Big\{ A^{+}_{\omega}(n) e^{-i\left[\omega t - k n \right]} \\
& \hspace{43mm} + A^{-}_{\omega}(n) e^{-i\left[ \omega t + k n \right]} \Big\} \nonumber 
\end{align}
with the slow varying envelopes $A^{\pm}_{\omega}(n)$ and with $k(\omega)$ as defined in Eq.~\eqref{eq:SignalWavenumber_Appendix}. Substituting the bare mode expansion from Eq.~\eqref{eq:BareModeExpansion_Appendix} into Eq.~\eqref{eq:EoM_m(x,t)_Appendix}, the dynamics of the system read
\begin{align}
   &\int_{-\infty}^{\infty} \mathrm{d}\omega \ \frac{\sqrt{Z_0}}{\sqrt{|\omega|}} \sum_{\sigma=\pm1} e^{-i[\omega t - \sigma k n]}\Bigg\{ \frac{-\omega^2}{\nu^2 }A^{\sigma}_{\omega}(n) \\
   & + \frac{\omega_{\mathrm{L}}^2}{\nu^2} A^{\sigma}_{\omega}(n) - \Big(1- \frac{\omega^2}{\omega_{\mathrm{J}}^2} \Big)\Big( A^{\sigma}_{\omega}(n+1)e^{\sigma ik} \nonumber - 2 A^{\sigma}_{\omega}(n) \nonumber \\
   &+ A^{\sigma}_{\omega}(n-1)e^{-\sigma ik} \Big) \Bigg\} 
   = \int_{-\infty}^{\infty} \mathrm{d}\omega' \ \frac{\sqrt{Z_0}}{\sqrt{|\omega'|}} \sum_{\sigma=\pm1} \nonumber \\
   & \Bigg\{
   m(n,t) \bigg( A^{\sigma}_{\omega'}(n+1)e^{\sigma ik'} -A^{\sigma}_{\omega'}(n) \bigg)
  - m(n-1,t) \nonumber \\
  & \hspace{8mm} \times \left(A^{\sigma}_{\omega'}(n) - A^{\sigma}_{\omega'}(n-1)e^{-\sigma ik'}\right) \Bigg\} e^{-i\left[ \omega' t -\sigma k' n \right]} \nonumber
\end{align}

Moreover, the envelopes are assumed to be varying on length scales $l$ much larger than a unit cell, i.e. $l\gg a$. In that limit 
\begin{equation}
  \left| \frac{\mathrm{d}^2A^{\pm}_{\omega}}{\mathrm{d}x^2}\right| \ll \left| \frac{\mathrm{d}A^{\pm}_{\omega}}{\mathrm{d}x}\right| \ll \left| A^{\pm}_{\omega}\right| 
\end{equation}

so that the higher order spatial variations can be neglected and we approximate
\begin{equation}
  A^{\pm}_{\omega}(n\pm 1) \approx A^{\pm}_{\omega}(n) \pm \frac{\mathrm{d}A^{\pm}_{\omega}(n)}{\mathrm{d}x} 
\end{equation}
In the Slow-Varying Envelope Approximation (SVEA), keeping the leading order terms \footnote{The first order derivative is conserved on the left-hand side as the other terms cancel out.}, the equations of motion read
\begin{align}
   &\int_{-\infty}^{\infty} \mathrm{d}\omega\ \frac{\sqrt{Z_0}}{\sqrt{|\omega|}} \sum_{\sigma=\pm 1} e^{-i[\omega t - \sigma kn]}\Bigg\{ \frac{-\omega^2}{\nu^2 }A^{\sigma}_{\omega} \\
   & + \frac{\omega_{\mathrm{L}}^2}{\nu^2} A^{\sigma}_{\omega} -2\Big(1- \frac{\omega^2}{\omega_{\mathrm{J}}^2}\Big)\Big[\big(\cos k-1\big) A^{\sigma}_{\omega} +\sigma i\sin k \frac{\mathrm{d} A^{\sigma}_{\omega}}{\mathrm{d}x} \Big] \Bigg\} \nonumber \\
   & = \int_{-\infty}^{\infty} \mathrm{d}\omega'\ \frac{\sqrt{Z_0}}{\sqrt{|\omega'|}} \sum_{\sigma=\pm1} 2i\sigma\sin\left(\frac{k'}{2}\right) \Bigg\{ m(n,t) e^{\sigma i\frac{k'}{2}} \nonumber \\
   & \hspace{ 23mm} - m(n-1,t) e^{-\sigma i\frac{k'}{2}} \Bigg\} A^{\sigma}_{\omega'} e^{-i\left[ \omega' t - \sigma k' n \right]} \nonumber
\end{align}
Using the dispersion relation from Eq.~\eqref{eq:SignalWavenumber_Appendix}, the equations of motion in the bare-mode basis simplify to 
\begin{align}
  & \int_{-\infty}^{\infty} \mathrm{d}\omega\ \frac{\sqrt{Z_0}}{\sqrt{|\omega|}}
   \Big(1- \frac{\omega^2}{\omega_{\mathrm{J}}^2}\Big) \sin k \sum_{\sigma=\pm 1} \sigma \frac{ \mathrm{d}A^{\sigma}_{\omega}}{\mathrm{d}x} \\
   &\times e^{-i[\omega t -\sigma kn]} = - \int_{-\infty}^{\infty} \mathrm{d}\omega'\ \frac{\sqrt{Z_0}}{\sqrt{|\omega'|}} \sin\left(\frac{k'}{2}\right) \sum_{\sigma=\pm 1}\sigma \nonumber \\
   & \times \Bigg\{
   m(n,t) e^{\sigma i\frac{k'}{2}} - m(n-1,t) e^{-\sigma i\frac{k'}{2}} \Bigg\} A^{\sigma}_{\omega'} e^{-i\left[ \omega' t -\sigma k' n \right]} \nonumber
\end{align}

\subsubsection*{Adiabatic Parametric Modulations}
We now consider the case of adiabatic parametric modulations. In that case, the modulation amplitude and phase-velocity varies slowly -- in a quasi-adiabatic fashion -- from cell to cell. In this adiabatic limit, we can assume that the variation of the modulation amplitude is negligibly small between two adjacent cells, that is
\begin{equation}
  |m(n)| \approx |m(n-1)|
\end{equation}
We can also assume that the wavenumber is approximately constant between adjacent cells, so
\begin{subequations}
  \begin{equation}
  \bar{\theta}_{\mathrm{p}}(n) \approx \theta_{\mathrm{p}}(n) + k_{\mathrm{p}}(n)/2
  \end{equation}
  \begin{equation}
  \bar{\theta}_{\mathrm{p}}(n-1) \approx \theta_{\mathrm{p}}(n) - k_{\mathrm{p}}(n)/2
  \end{equation}
\end{subequations}
where $\theta(n) = \sum_{x=0}^n k_{\mathrm{p}}(n)$. Then, the equations of motion with this adiabatic modulation read
\begin{align}
   &\int_{-\infty}^{\infty} \mathrm{d}\omega\ \frac{\sqrt{Z_0}}{\sqrt{|\omega|}}
   \Big(1- \frac{\omega^2}{\omega_{\mathrm{J}}^2}\Big)\sin k \sum_{\sigma=\pm 1} \sigma \frac{\mathrm{d} A^{\sigma}_{\omega}}{\mathrm{d}x} \\
   & \times e^{-i[\omega t -\sigma kn]} = |m(n)| \int_{-\infty}^{\infty} \mathrm{d}\omega' \frac{\sqrt{Z_0}}{\sqrt{|\omega'|}} \sin\left(\frac{k'}{2}\right) \nonumber \nonumber \\
   & \times \sum_{\sigma=\pm 1} \sigma A^{\sigma}_{\omega'} e^{-i\left[ \omega' t -\sigma k' n \right]} \Bigg\{ \sin(\omega_{\mathrm{p}} t- \theta_{\mathrm{p}}-\frac{k_{\mathrm{p}}}{2}) e^{\sigma i\frac{k'}{2}} \nonumber \\
   &\hspace{39mm} - \sin(\omega_{\mathrm{p}} t- \theta_{\mathrm{p}}+\frac{k_{\mathrm{p}}}{2}) e^{-\sigma i\frac{k'}{2}} \Bigg\} \nonumber
\end{align}
and, after some trigonometric manipulations, using the fact that 
\begin{align}
  &\sin(A-B)e^{iC} - \sin(A+B)e^{-iC} = \\
  &\frac{1}{2i} \big(e^{iA}e^{-iB}e^{iC} - e^{-iA}e^{iB}e^{iC} - e^{iA}e^{iB}e^{-iC}+ e^{-iA}e^{-iB}e^{-iC}\big) \nonumber \\
  & \hspace{0mm} = -2 \Big[ \cos A \sin B \cos C - i\sin A \cos B \sin C\Big] \nonumber
\end{align}
the equation of motion can be rewritten as
\begin{align}
   &\int_{-\infty}^{\infty} \mathrm{d}\omega\ \frac{\sqrt{Z_0}}{\sqrt{|\omega|}}
   \Big(1- \frac{\omega^2}{\omega_{\mathrm{J}}^2}\Big)\sin k \sum_{\sigma=\pm1} \sigma\frac{\mathrm{d} A^{\sigma}_{\omega}}{\mathrm{d}x} \\ & \times e^{-i[\omega t -\sigma kn]} = - |m| \int_{-\infty}^{\infty}\mathrm{d}\omega'\ \frac{\sqrt{Z_0}}{\sqrt{|\omega'|}} \sin\left(\frac{k'}{2}\right) \nonumber \\
   & \times \sum_{\sigma=\pm1} \sigma A^{\sigma}_{\omega'} e^{-i\left[ \omega' t -\sigma k' n \right]} 2\Bigg\{ \cos(\omega_{\mathrm{p}} t- \theta_{\mathrm{p}}) \sin\Big(\frac{k_{\mathrm{p}}}{2}\Big) \nonumber \\
   & \hspace{5mm} \times\cos \Big(\frac{k'}{2}\Big) - i \sigma \sin(\omega_{\mathrm{p}} t- \theta_{\mathrm{p}}) \cos\Big(\frac{k_{\mathrm{p}}}{2} \Big) \sin \Big( \frac{k'}{2} \Big) \Bigg\} \nonumber 
\end{align}
where we used $|m|=|m(n)|$ for ease of notation. The $\sin$ and $\cos$ can be expressed as sums of complex exponentials
\begin{align}
   &\int_{-\infty}^{\infty} \mathrm{d}\omega\ \frac{\sqrt{Z_0}}{\sqrt{|\omega|}}
   \Big(1- \frac{\omega^2}{\omega_{\mathrm{J}}^2}\Big)\sin k \sum_{\sigma=\pm1} \sigma \frac{\mathrm{d} A^{\sigma}_{\omega}}{\mathrm{d}x} \\
   &\times e^{-i[\omega t -\sigma kn]} =  -|m| \int_{-\infty}^{\infty} \mathrm{d}\omega'\ \frac{\sqrt{Z_0}}{\sqrt{|\omega'|}} \sin\left(\frac{k'}{2}\right) \nonumber \\ & \times \sum_{\sigma=\pm1} \sigma A^{\sigma}_{\omega'} e^{-i\left[ \omega' t -\sigma k' n \right]} \Bigg\{
   \sin\Big(\frac{k_{\mathrm{p}}}{2}\Big) \cos \Big(\frac{k'}{2}\Big)  e^{i[\omega_{\mathrm{p}} t- \theta_{\mathrm{p}}]} \nonumber \\
   & \hspace{39mm} + \sin\Big(\frac{k_{\mathrm{p}}}{2}\Big) \cos \Big(\frac{k'}{2}\Big) e^{-i[\omega_{\mathrm{p}} t- \theta_{\mathrm{p}}]} \nonumber \\
   &\hspace{39mm}  -\sigma \cos\Big(\frac{k_{\mathrm{p}}}{2} \Big) \sin \Big( \frac{k'}{2} \Big) e^{i[\omega_{\mathrm{p}} t- \theta_{\mathrm{p}}]} \nonumber\\
   & \hspace{39mm} +\sigma \cos\Big(\frac{k_{\mathrm{p}}}{2} \Big) \sin \Big( \frac{k'}{2} \Big) e^{-i[\omega_{\mathrm{p}} t- \theta_{\mathrm{p}}]} \Bigg\} \nonumber
\end{align}
that we can simplify to, since $\sin(a \pm b) = \sin(a)\cos(b) \pm \cos(a)\sin(b)$,
\begin{align}
   &\int_{-\infty}^{\infty} \mathrm{d}\omega\ \frac{\sqrt{Z_0}}{\sqrt{|\omega|}}
   \Big(1- \frac{\omega^2}{\omega_{\mathrm{J}}^2}\Big)\sin k \sum_{\sigma=\pm1} \sigma \frac{\mathrm{d} A^{\sigma}_{\omega}}{\mathrm{d}x} \\
   & \times e^{-i[\omega t -\sigma kn]} = |m| \int_{-\infty}^{\infty}\mathrm{d}\omega'\ \frac{\sqrt{Z_0}}{\sqrt{|\omega'|}} \sin\left(\frac{k'}{2}\right) \nonumber \\
   & \times \sum_{\sigma=\pm1} \sigma^2 A^{\sigma}_{\omega'} e^{-i\left[ \omega' t -\sigma k' n \right]}
   \Bigg\{ \sin\Big(\frac{k'-\sigma k_{\mathrm{p}}}{2}\Big) e^{i[\omega_{\mathrm{p}} t- \theta_{\mathrm{p}}]} \nonumber \\
   & \hspace{40 mm} - \sin\Big(\frac{k'+\sigma k_{\mathrm{p}}}{2}\Big) e^{-i[\omega_{\mathrm{p}} t- \theta_{\mathrm{p}}]} \nonumber \Bigg\} 
\end{align}
and since $e^a e^b = e^{(a+b)}$, we rewrite
\begin{align}
   &\int_{-\infty}^{\infty} \mathrm{d}\omega\ \frac{\sqrt{Z_0}}{\sqrt{|\omega|}}
   \Big(1- \frac{\omega^2}{\omega_{\mathrm{J}}^2}\Big)\sin k \sum_{\sigma=\pm1} \sigma \frac{\mathrm{d} A^{\sigma}_{\omega}}{\mathrm{d}x} \\
   &\times e^{-i[\omega t -\sigma kn]} =|m| \int_{-\infty}^{\infty} \mathrm{d}\omega'\ \frac{\sqrt{Z_0}}{\sqrt{|\omega'|}} \sin\left(\frac{k'}{2}\right) \nonumber \\ 
   & \times \sum_{\sigma=\pm1} A^{\sigma}_{\omega'} \Bigg\{ \sin\Big(\frac{k'-\sigma k_{\mathrm{p}}}{2}\Big) e^{-i[\omega'-\omega_{\mathrm{p}}] t} e^{\sigma i [k'n - \sigma \theta_{\mathrm{p}}]} \nonumber \\
   & \hspace{17mm}- \sin\Big(\frac{k'+\sigma k_{\mathrm{p}}}{2}\Big) e^{-i[\omega'+\omega_{\mathrm{p}}] t} e^{\sigma i [k'n + \sigma \theta_{\mathrm{p}}]} \Bigg\} \nonumber 
\end{align}
\subsubsection*{Rotating Wave Approximation}
As seen in the previous section, the dynamics consist of an infinite number of terms that oscillate in time. However, considering a frame rotating as $e^{-i\omega t}$, the fast-oscillating terms in this frame quickly average to zero while the slower-oscillating terms contribute more significantly. In the Rotating-Wave Approximation (RWA), the fast oscillating terms are therefore neglected yielding
\begin{align}
   & \frac{\sqrt{Z_0(\omega)}}{\sqrt{|\omega|}}
   \Big(1- \frac{\omega^2}{\omega_{\mathrm{J}}^2}\Big)\sin k \sum_{\sigma=\pm1} \sigma \frac{\mathrm{d} A^{\sigma}_{\omega}}{\mathrm{d}x} e^{\sigma i kn} \label{eq:EoMTemporalRWA}\\
   & = |m| \sum_{\sigma=\pm1} \Bigg\{  \sin\Big(\frac{k_{\Sigma}-\sigma k_{\mathrm{p}}}{2}\Big) \sin\left(\frac{k_{\Sigma}}{2}\right) \frac{\sqrt{Z_0(\omega_\Sigma)}}{\sqrt{|\omega_{\Sigma}|}} \nonumber \\
   & \times A^{\sigma}_{\omega_{\Sigma}} e^{\sigma i[k_{\Sigma} n -\sigma \theta_{\mathrm{p}}]} - \sin\Big(\frac{k_{\Delta}+\sigma k_{\mathrm{p}}}{2}\Big) \sin\left(\frac{k_{\Delta}}{2}\right) \frac{\sqrt{Z_0(\omega_\Delta)}}{\sqrt{|\omega_{\Delta}|}} \nonumber \\
   & \hspace{55mm} \times A^{\sigma}_{\omega_{\Delta}} e^{\sigma i [k_{\Delta}n + \sigma \theta_{\mathrm{p}}]} \Bigg\} \nonumber
\end{align}
where $\omega_\Sigma = \omega+ \omega_{\mathrm{p}}$ and $\omega_\Delta = \omega- \omega_{\mathrm{p}}$ and we use the short-
hand notation $k_\Sigma = k(\omega_\Sigma)$ and $k_\Delta = k(\omega_\Delta)$. 
The sinusoidal flux-modulations with angular frequency $\omega_{\mathrm{p}}$ hence result in parametric coupling between the right- and left-propagating modes at frequencies $\omega$, $\omega_{\Sigma} = \omega+\omega_{\mathrm{p}}$ and $\omega_{\Delta} = \omega -\omega_{\mathrm{p}}$. \par
\subsection{Phase Matching}
The equation of motion in the temporal RWA, given in Eq. \eqref{eq:EoMTemporalRWA}, can be expanded as
\begin{align}
   &\frac{Z_0(\omega)}{\sqrt{|\omega|}}
   \big(1- \frac{\omega^2}{\omega_{\mathrm{J}}^2}\big) \sin k \Big(\frac{\mathrm{d} A^{+}_{\omega}}{\mathrm{d}x} e^{ikn} - \frac{\mathrm{d} A^{-}_{\omega}}{\mathrm{d}x} e^{-ikn}\Big)
   \\
   & = |m| \Bigg\{  \sin\Big(\frac{k_{\Sigma}-k_{\mathrm{p}}}{2}\Big) \sin\left(\frac{k_{\Sigma}}{2}\right) \frac{\sqrt{Z_0(\omega_{\Sigma})}}{\sqrt{|\omega_{\Sigma}|}} A^{+}_{\omega_{\Sigma}} e^{i\left[k_{\Sigma} n- \theta_{\mathrm{p}}\right]} \nonumber \\
   &\hspace{10mm}  - \sin\Big(\frac{k_{\Delta}+k_{\mathrm{p}}}{2}\Big) \sin\left(\frac{k_{\Delta}}{2}\right) \frac{\sqrt{Z_0(\omega_{\Delta})}}{\sqrt{|\omega_{\Delta}|}} A^{+}_{\omega_{\Delta}} e^{ i \left[k_{\Delta}n + \theta_{\mathrm{p}} \right]} \nonumber \\
   & \hspace{10mm} + \sin\Big(\frac{k_{\Sigma}+k_{\mathrm{p}}}{2}\Big) \sin\left(\frac{k_{\Sigma}}{2}\right) \frac{\sqrt{Z_0(\omega_{\Sigma})}}{\sqrt{|\omega_{\Sigma}|}} A^{-}_{\omega_{\Sigma}} e^{-i \left[k_{\Sigma} n+ \theta_{\mathrm{p}} \right]} \nonumber \\
   & \hspace{10mm} - \sin\Big(\frac{k_{\Delta}-k_{\mathrm{p}}}{2}\Big) \sin\left(\frac{k_{\Delta}}{2}\right) \frac{\sqrt{Z_0(\omega_{\Delta})}}{\sqrt{|\omega_{\Delta}|}} A^{-}_{\omega_{\Delta}} e^{- i \left[k_{\Delta} n - \theta_{\mathrm{p}} \right]} \Bigg\} \nonumber
\end{align}
Here too, the rapidly-oscillating terms quickly average to zero while the slowly varying terms dominate the dynamics of the system.
For a right-handed transmission line ($\nu_g = \partial \omega/\partial k \geq 0$) and a right-propagating pump ( $k_{\mathrm{p}} = \partial \theta_{\mathrm{p}}/\partial x>0$), we know that
\begin{subequations}
  \begin{equation}
  \left|k_{\Sigma}-k - k_{\mathrm{p}}\right| \ll \left|k_{\Sigma}+k + k_{\mathrm{p}}\right|
  \end{equation}
  \begin{equation}
  \left|k_{\Delta}-k + k_{\mathrm{p}}\right| \ll \left|k_{\Delta}+k - k_{\mathrm{p}} \right|
  \end{equation}
\end{subequations}
In the spatial RWA, the fast oscillating terms are hence neglected and the right-propagating modes fully decouple from the left-propagating ones, yielding
\begin{align}
   &\frac{\sqrt{Z_0(\omega)}}{\sqrt{|\omega|}}
   \Big(1- \frac{\omega^2}{\omega_{\mathrm{J}}^2}\Big)\sin k \frac{\mathrm{d} A^{\pm}_{\omega}}{\mathrm{d}x} = \pm |m| \\
   &\times \Bigg\{\sin\Big(\frac{k_{\Sigma}\mp k_{\mathrm{p}}}{2}\Big) \sin\left(\frac{k_{\Sigma}}{2}\right) \frac{\sqrt{Z_0(\omega_{\Sigma})}}{\sqrt{|\omega_{\Sigma}|}} A^{\pm}_{\omega_{\Sigma}} e^{\pm i\Delta \theta^\pm_\Sigma} \nonumber \\
   & \hspace{4mm} - \sin\Big(\frac{k_{\Delta}\pm k_{\mathrm{p}}}{2}\Big) \sin\left(\frac{k_{\Delta}}{2}\right) \frac{\sqrt{Z_0(\omega_{\Delta})}}{\sqrt{|\omega_{\Delta}|}} A^{\pm}_{\omega_{\Delta}} e^{ \pm i \Delta \theta^\pm_\Delta} \Bigg\} \nonumber
\end{align}
where we defined the accumulated phase mismatch as
\begin{subequations}
\label{eq:PhaseMismatch_Appendix}
  \begin{equation}
  \Delta \theta^{\pm}_{\Sigma} = \sum_0^n \Delta k^\pm_{\Sigma}\ = \sum_0^n (k_{\Sigma}-k \mp k_{\mathrm{p}}) 
  \end{equation}
  \begin{equation} 
  \Delta \theta^{\pm}_{\Delta} = \sum_0^n \Delta k^\pm_{\Delta}\ = \sum_0^n (k_{\Delta}-k \pm k_{\mathrm{p}}) 
  \end{equation}
\end{subequations}
This mismatch $\Delta k^\pm_{\Sigma/\Delta}$ can be seen in a dispersion diagrams. As shown in Fig.~\ref{fig:SystemDescription}b, phase matching is satisfied when the temporal and spatial angular frequencies of the modulation, i.e. $\omega_{\mathrm{p}}$ and $k_{\mathrm{p}}$, respectively, match the differences in temporal and spatial angular frequencies between the two coupled modes, e.g. between the two modes at $\omega_{\mathrm{s}}$ and $\omega_\Sigma$. In that case, the two modes are strongly coupled. When the modulation is not phase-matched, the contribution of the fast-oscillatory terms is negligibly small and leads to very weak parametric coupling between the modes.\par

This phase matching condition cannot simultaneously be satisfied in both direction of propagation as can be seen in Fig.~\ref{fig:SystemDescription}b. Considering a right-propagating sinusoidal modulation with $k_{\mathrm{p}}>0$, we see in Eq.~\eqref{eq:PhaseMismatch_Appendix} that if $\Delta k_{\Sigma}^{\pm} = 0$ then $\Delta k_{\Sigma}^{\mp} \neq 0$.
The time-reversal symmetry of the system is hence broken leading to a direction-dependent parametric coupling between the modes.
Note however that only spatio-temporal modulations can produce such non-reciprocity. Transmission lines with purely spatial ($\omega_{\mathrm{p}}=0$) or purely temporal ($k_{\mathrm{p}}=0$) parametric modulations are fully reciprocal and present the same parametric coupling irrespective of the direction of propagation.\par
 
The equations of motion for this system can hence be expressed as a set of coupled modes equations
\begin{align}
   &\frac{\mathrm{d} A^{\pm}_{\omega}}{\mathrm{d}x} = \frac{|m|}{\sin(k)(1-\omega^2/\omega_{\mathrm{J}}^2)}\\
   \times \Bigg\{&\sin\Big(\frac{k_{\Sigma}\mp k_{\mathrm{p}}}{2}\Big) \sin\left(\frac{k_{\Sigma}}{2}\right) \frac{\sqrt{|\omega|}}{\sqrt{|\omega_{\Sigma}|}} \frac{\sqrt{Z_0(\omega_{\Sigma})}}{\sqrt{Z_0(\omega)}} A^{\pm}_{\omega_{\Sigma}} e^{\pm i\Delta \theta^\pm_\Sigma} \nonumber \\
   - &\sin\Big(\frac{k_{\Delta}\pm k_{\mathrm{p}}}{2}\Big)
  \sin\left(\frac{k_{\Delta}}{2}\right) \frac{\sqrt{|\omega|}}{\sqrt{|\omega_{\Delta}|}} \frac{\sqrt{Z_0(\omega_{\Delta})}}{\sqrt{Z_0(\omega)}} A^{\pm}_{\omega_{\Delta}} e^{\pm i \Delta \theta^\pm_\Delta} \Bigg\} \nonumber
\end{align}
For a subset of two right-propagating modes with frequencies $\omega$ and $\omega_\Sigma =\omega+\omega_{\mathrm{p}}$, the dynamics in this two-level system reads
\begin{equation}
  \frac{\mathrm{d}}{\mathrm{d}x} \begin{bmatrix} A^{+}_{\omega_{\Sigma}} \\ A^{+}_{\omega}
  \end{bmatrix}
  = \begin{bmatrix} 0 & -a e^{-i\Delta \theta_\Sigma^+} \\ b e^{i\Delta \theta_\Sigma^+}& 0 
\end{bmatrix}
\cdot 
\begin{bmatrix} A^{+}_{\omega_{\Sigma}} \\ A^{+}_{\omega} \end{bmatrix}
\end{equation}
with the coupling coefficients 
\begin{subequations}\label{eq:coeffs_Appendix}
  \begin{equation}
  a =|m| \frac{\sqrt{|\omega_\Sigma|\ Z_0(\omega)}}{\sqrt{|\omega|\ Z_0(\omega_\Sigma)}} \frac{\sin\left(\frac{k}{2}\right) }{\sin(k_\Sigma)} \frac{\sin\left(\frac{k+k_{\mathrm{p}}}{2}\right) }{(1-\omega_{\Sigma}^2/\omega_{\mathrm{J}}^2)} 
  \end{equation}
  \begin{equation}
  b = |m| \frac{\sqrt{|\omega|\ Z_0(\omega_{\Sigma})}}{\sqrt{|\omega_{\Sigma}|\ Z_0(\omega)}} \frac{\sin\left(\frac{k_{\Sigma}}{2}\right) }{\sin(k)} \frac{\sin\left(\frac{k_{\Sigma}-k_{\mathrm{p}}}{2}\right) }{\left(1-\omega^2/\omega_{\mathrm{J}}^2\right)}
  \end{equation}
\end{subequations}
Note that these two coefficients are equal when the modulation is phase-matched, i.e. $\Delta k^+_\Sigma=0$. Therefore, they can be approximated as $a=b=g$ whenever $\Delta k^+_\Sigma \ll k$. 

\section{Superconducting Quantum Interference Device}\label{appendix:SQUID}
\begin{figure}[ht]
  \centering
  \includegraphics[width=0.4\linewidth]{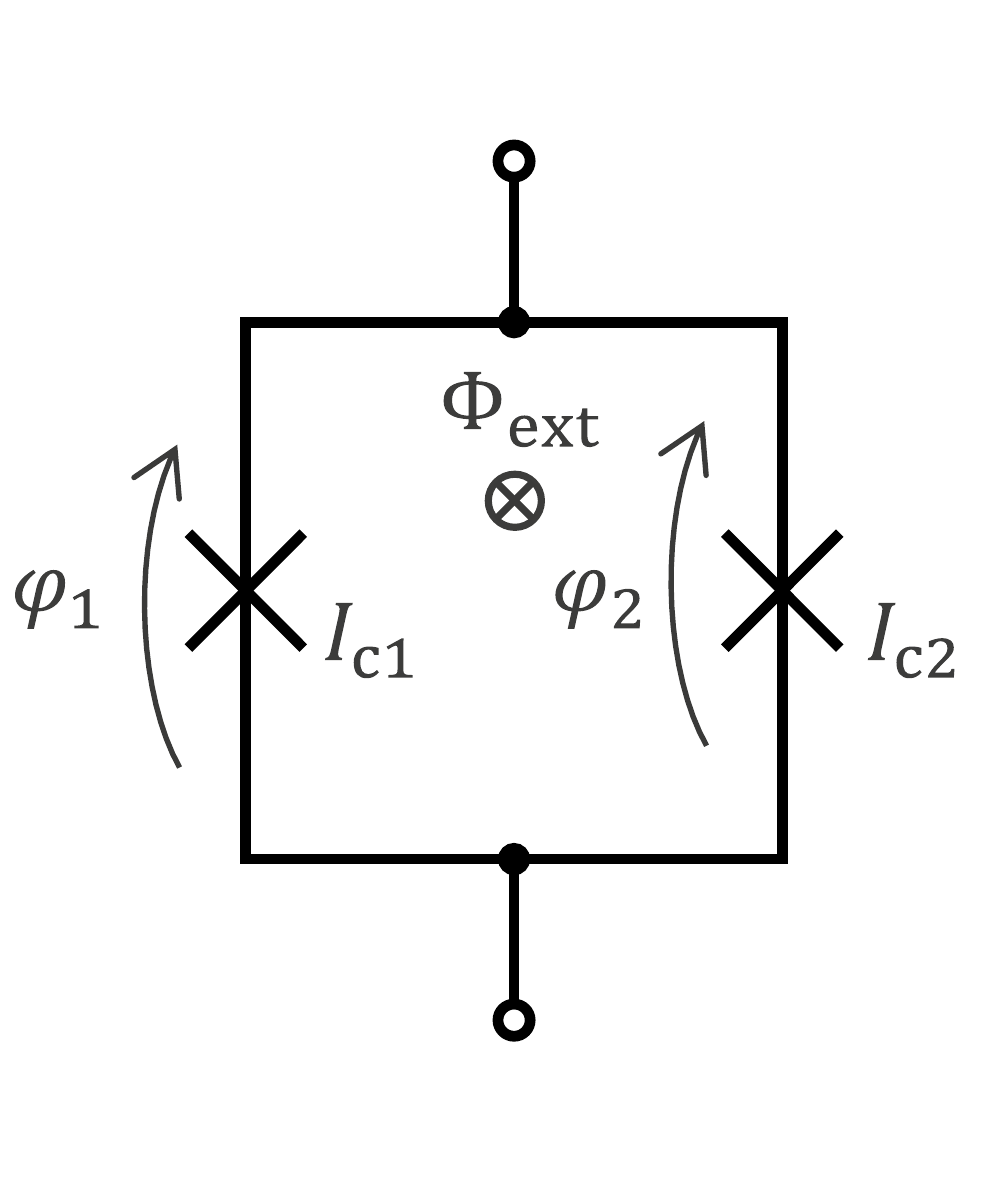}
  \caption{\textbf{Direct current Superconducting Quantum Interference Device}. The circuit nicknamed \textit{DC-SQUID} consists of two Josephson junctions in parallel, with critical current $I_{\mathrm{c}1}$ and $I_{\mathrm{c}2}$, forming a loop threaded by an external flux $\Phi_{\mathrm{ext}}$. }
  \label{fig:SQUID}
\end{figure}
Consider a SQUID, formed by two junctions in parallel forming a superconducting loop, as depicted in Fig.~\ref{fig:SQUID}. Due to the flux-quantization around the loop, the normalized phase difference across the two junctions, $\varphi_{1/2}$, and the external magnetic flux threading the SQUID must satisfy the following relation $\varphi_1 +\Phi_{\mathrm{ext}}/\phi_0 = \varphi_2 \ (\mathrm{mod} \ 2\pi)$ \cite{Vool2017}. The potential energy of the SQUID is then found as
\begin{align}
  U_{\mathrm{SQUID}} = &-E_{\mathrm{J}1} \cos\varphi_1 - E_{\mathrm{J}2} \cos \varphi_2\\
  = &-\left(\frac{E_{\mathrm{J}1}+E_{\mathrm{J}2}}{2}\right)\left( \cos \varphi_1 + \cos \varphi_2 \right) \nonumber \\
  &  -\left(\frac{E_{\mathrm{J}1}-E_{\mathrm{J}2}}{2}\right)\left( \cos \varphi_1 - \cos \varphi_2 \right) \nonumber\\
  =& -\left( E_{\mathrm{J}1} + E_{\mathrm{J}2}\right) \cos\left( \frac{\varphi_1+\varphi_2}{2}\right) \cos\left( \frac{\varphi_1-\varphi_2}{2}\right) \nonumber\\
  &  -\left( E_{\mathrm{J}1} - E_{\mathrm{J}2}\right) \sin\left( \frac{\varphi_1+\varphi_2}{2}\right) \sin\left( \frac{\varphi_1-\varphi_2}{2}\right) \nonumber
\end{align}
with $E_{\mathrm{J}i} = \phi_0 I_{\mathrm{c}i}$. Replacing $\varphi_1-\varphi_2 = -\Phi_{\mathrm{ext}}/\phi_0$, we can re-write 
\begin{align}
  U_{\mathrm{SQUID}} = -&E_{J\Sigma} \cos\left( \frac{\Phi_{\mathrm{ext}}}{2\phi_0}\right)\Bigg[ \cos\left( \frac{\varphi_1 + \varphi_2}{2}\right) \\
  &  + d \tan \left( \frac{\Phi_{\mathrm{ext}}}{2\phi_0} \right) \sin\left( \frac{\varphi_1 + \varphi_2}{2}\right)\Bigg] \nonumber
\end{align}
with $d = (E_{\mathrm{J}1}-E_{\mathrm{J}2})/(E_{\mathrm{J}1}+E_{\mathrm{J}2})$ and $E_{J\Sigma}=(E_{\mathrm{J}1}+E_{\mathrm{J}2})$. For ease of notation, let's introduce $\xi$, so that $\tan \xi = d \tan \left( \frac{\Phi_{\mathrm{ext}}}{2\phi_0} \right)$ and $1/\cos\xi = \sqrt{1+\tan \xi}$. We can then re-write
\begin{align}
  & U_{\mathrm{SQUID}} =-E_{J\Sigma} \frac{\cos\left( \frac{\Phi_{\mathrm{ext}}}{2\phi_0}\right)}{\cos \xi}\Bigg[ \cos\left( \frac{\varphi_1 + \varphi_2}{2}\right) \\
  & \times \cos\xi + \sin\left( \frac{\varphi_1 + \varphi_2}{2}\right) \sin \xi \Bigg] = -E_{J\Sigma} \cos\left( \frac{\Phi_{\mathrm{ext}}}{2\phi_0}\right) \nonumber \\
  & \hspace{13mm} \times \sqrt{1+d^2\tan^2\left(\frac{\Phi_{\mathrm{ext}}}{2\phi_0}\right)} \cos\left(\frac{\varphi_1+\varphi_2}{2}-\xi \right) \nonumber
\end{align}
This is the same expression as in Ref. \cite{Vool2017} in eq.(4.19). In the symmetric DC-SQUIDs with $E_{\mathrm{J}1}=E_{\mathrm{J}2}=E_{\mathrm{J}0}$ and $\varphi_1=\varphi_2 = \varphi $, the coefficient $d=0$ such that $\xi=0\ (\mathrm{mod}\ \pi)$
and the DC-SQUID's energy simplifies to 
\begin{equation}
  U_{\mathrm{SQUID}} =-2E_{\mathrm{J}0} \cos\left( \frac{\Phi_{\mathrm{ext}}}{2\phi_0}\right)\cos\varphi
\end{equation}
So the potential energy of the DC-SQUID is equivalent to the potential energy of a single junction with a flux-tunable energy $E_{\mathrm{J},\mathrm{SQUID}}(\Phi_{\mathrm{ext}})= 2E_{\mathrm{J}0} \cos\left( \Phi_{\mathrm{ext}}/2\phi_0\right) $

\section{Flux-Modulated SQUID Model} \label{appendix:ADSSQUIDModel}
The proposed circuit shown in Fig.~\ref{fig:circuit} includes flux-modulated SQUIDs that are not readily available in ADS Keysight \cite{Shiri2024,Naaman2025}. To simulate and evaluate the performance of the proposed circuit with the HB method, we have developed a model for flux-modulated SQUIDs, described in the following. \par
The dynamics of a JJ can be expressed in terms of the current $I(t)$ through and voltage $V(t)$ across the JJ with the following Josephson relations
\begin{align}
 I(t) = I_{\mathrm{c}} \sin(\varphi)\label{eq:JJCurrent_Appendix}
 \hspace{5mm} \mathrm{ and } \hspace{5mm} 
\varphi(t) &= \frac{1}{\phi_0}\int_{-\infty}^t V(\tau) \mathrm{d} \tau
\end{align}
with the critical current of the junction $I_{\mathrm{c}}$ \cite{Vool2017}.
In these relations, the junction is assumed to operate in its superconducting state where it is lossless, i.e. $I<I_{\mathrm{c}}$ and $T=0\ K$, so that $R_{\mathrm{N}}=0$. \par
Considering a DC-SQUID (Appendix \ref{appendix:SQUID}), quantization of the magnetic flux imposes the additional relation between the flux $\varphi_{1/2}$ across the two junctions
\begin{equation}\label{eq:FluxQuantization_Appendix}
\varphi_1 + \Phi_{\mathrm{ext}}/\phi_0 = \varphi_2 \ (\mathrm{mod} \ 2\pi)
\end{equation}
where $\Phi_{\mathrm{ext}}$ is the flux resulting from the external magnetic field threading the SQUID, as depicted in Fig.~\ref{fig:SQUID}.\par
\begin{figure}[ht]
  \centering
  \includegraphics[width=1\linewidth]{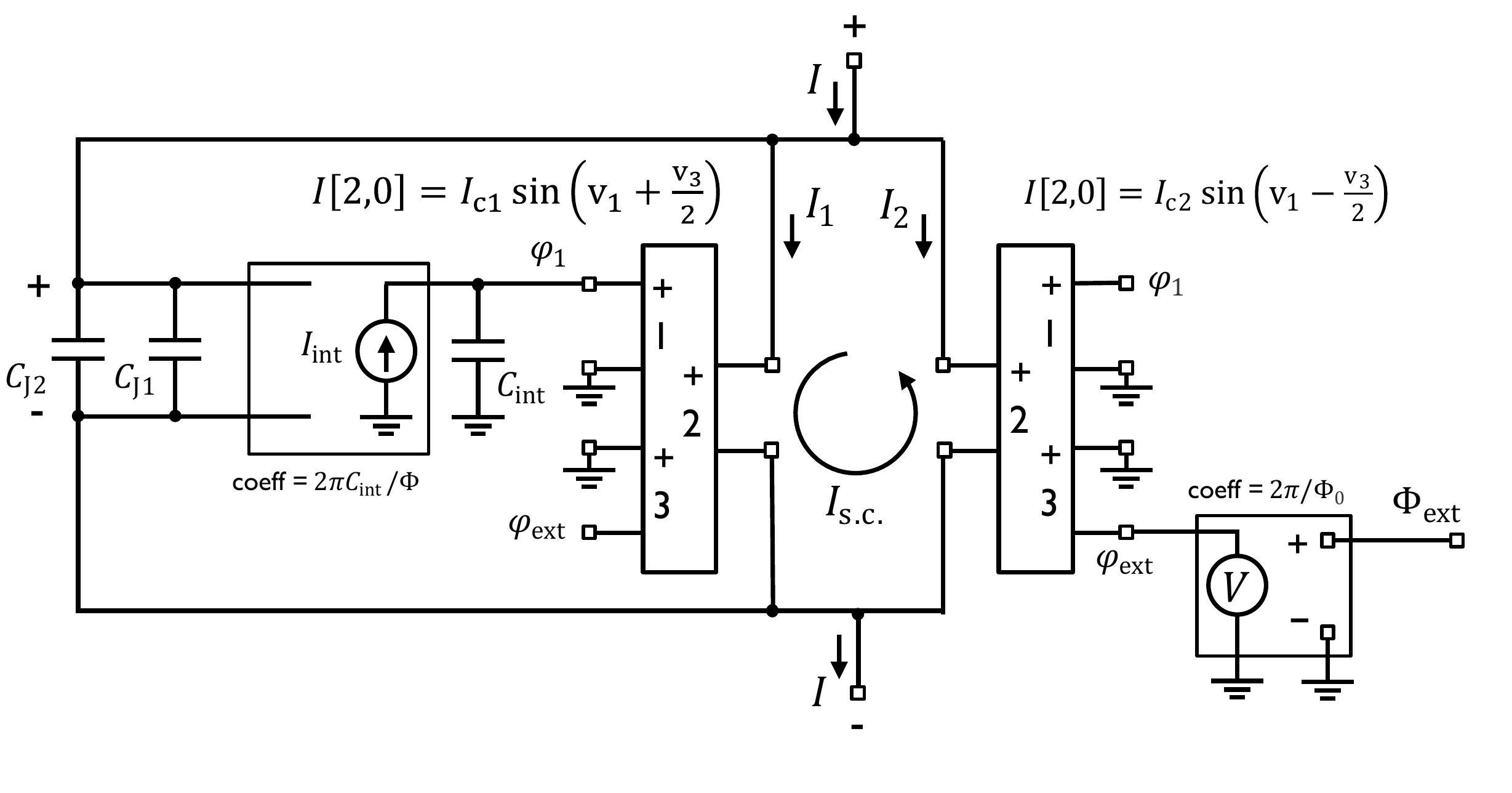}
  \caption{\textbf{Circuit diagram of the ADS model for the DC-SQUID}. The current through the two branches of the SQUID are specified as $I_{1/2} = I_{\mathrm{c}1/2}\sin(\varphi_1\pm 0.5\Phi_{\mathrm{ext}}/\phi_0)$ where $\varphi_1$ i in parallel, with critical current $I_{\mathrm{c}1}$ and $I_{\mathrm{c}2}$, forming a loop threaded by an external flux $\Phi_{\mathrm{ext}}$ that induces a superconducting circulating current $I_{\mathrm{s.c.}}$. }
  \label{fig:ADSModelSQUID_Appendix}
\end{figure}
The dynamics given by Eqs.~\eqref{eq:JJCurrent_Appendix} and \eqref{eq:FluxQuantization_Appendix} are captured in the SQUID-model presented in Fig.~\ref{fig:ADSModelSQUID_Appendix}. Similar to the approach presented in Ref. \cite{Shiri2024}, the phase across the junction from Eq.~\eqref{eq:JJCurrent_Appendix} is obtained with an integrator formed by a voltage-controlled current source charging a capacitor, i.e. 
\begin{equation}
  \varphi_1 = \frac{1}{C_{\mathrm{int}}}\int_0^t I_{\mathrm{int}}(\tau) \mathrm{d}\tau = \frac{1}{\phi_0}\int_0^t V(\tau) \mathrm{d}\tau
\end{equation}
Given the phase $\varphi_1$, Symbolic Defined Devices (SDDs) are then used to force the right current in each branch of the SQUID.\par

The SDDs are n-port devices where the currents at each port is specified as a function of the other ports' voltages and currents, i.e. $i_k =f(\{v_k,i_k\})$. The two JJs in the SQUID are thus realized with 3-port SDDs. Port $1$ of each SDD is connected to the integrator output $\varphi_1$ and their current set to $i_1=0$ to guarantee an infinite input impedance. Similarly, port $3$ of each junction has infinite input impedance and is connected to $\varphi _{\mathrm{ext}}=\Phi_{\mathrm{ext}}/\phi_0$. To model the Josephson current from Eq.~\eqref{eq:JJCurrent_Appendix}, the current through port $2$ of the SDDs is specified as $i_{1/2} = I_{\mathrm{c}1/2} \sin(v_1\pm v_3/2)$ which includes the flux-quantization from Eq.~\eqref{eq:FluxQuantization_Appendix}. Moreover, the parasitic capacitance of the two junctions is modeled with two capacitors $C_{\mathrm{J}1}$ and $C_{\mathrm{J}2}$ in parallel with the SDDs. \par
With this model for the SQUID, flux-modulation must be imposed. Contrary to Fig.~\ref{fig:circuit}, the model used in ADS Keysight for simulation does not include the inductive coupling to a pump line. Instead, the external fluxes threading the SQUIDs in every unit cell was modeled by AC voltage sources connected to $\Phi_{\mathrm{ext}}$ with amplitude and phase that varies depending on the unit-cell number. \par
\begin{figure}
  \centering
  \includegraphics[width=0.85\linewidth]{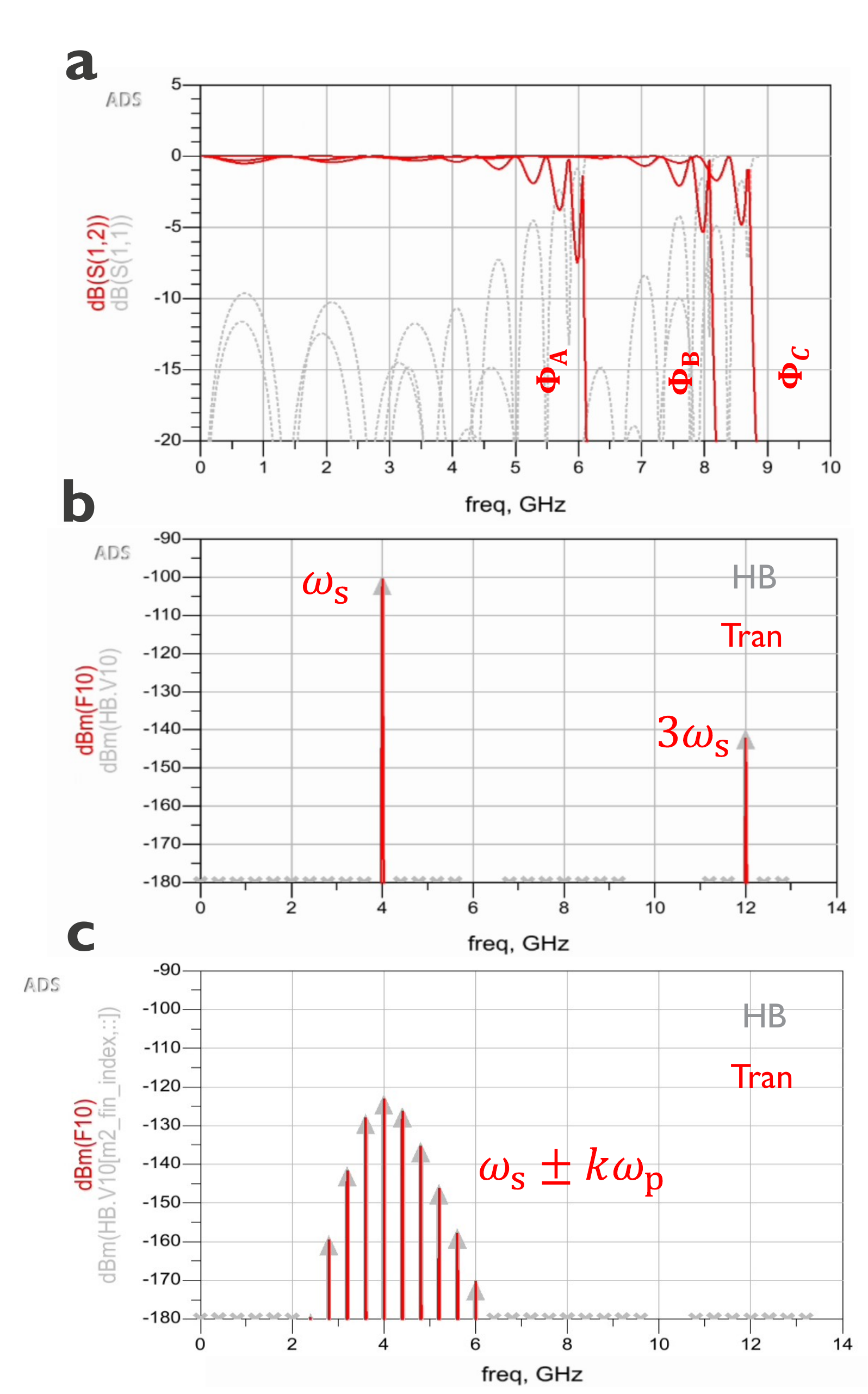}
  \caption{\textbf{Simulated artificial transmission line} with our flux-modulated SQUID model \textbf{(a)} 
  In SP simulations, the Bragg frequency of the line varies with the applied DC flux bias $\Phi_{A/B/C}$ which controls the SQUID's equivalent inductance. 
  \textbf{(b)} 
  In HB and Tran simulation, harmonic generation is observed due to the SQUID's nonlinearity. 
  \textbf{(c)} In HB and Tran simulations, AC flux modulations result in parametric coupling between the modes at $\omega_{\mathrm{s}}\pm k\omega_{\mathrm{p}}$, with $k\in \mathbb{Z}$.}
  \label{fig:SQUID_TL_SimulationVertical}
\end{figure}

As shown in Fig.~\ref{fig:SQUID_TL_SimulationVertical}, this model for flux-modulated SQUID can be used in ADS Keysight for S-parameter (SP), Transient (Tran) and Harmonic Balance (HB) simulations. S-parameter simulations with the model could capture the inductance variation of the SQUID, and therefore change in Bragg frequency of the line, as a function of the external flux (Fig.~\ref{fig:SQUID_TL_SimulationVertical}a). The model also capture the effect of the Junctions' non-linearities in both Tran and HB simulations where third harmonic appeared for large input signals (Fig.~\ref{fig:SQUID_TL_SimulationVertical}b). Finally, the proposed model produced the modulation-induced parametric coupling, both in Tran and HB simulation. As seen in Fig.~\ref{fig:SQUID_TL_SimulationVertical}c, flux-modulations with frequency $\omega_{\mathrm{p}}$ redistributes the photons propagating initially at $\omega_{\mathrm{s}}$ to other modes propagating at $\omega_{\mathrm{s}}\pm k \omega_{\mathrm{p}}$, $k\in \mathbb{Z}$.
Note that although both Tran and HB simulation lead to the same results, HB is much faster than Tran analysis and speed up the simulation time by orders of magnitude \cite{Shiri2024}.
\newpage

\end{document}